%% file: ms.tex
\documentclass[12pt,preprint]{aastex}
\usepackage{lscape}

\newcommand{\rcav}{R_{\rm cav}}
\newcommand{\cs}{c_{\rm s}}
\newcommand{\dc}{\delta_{\rm cav}}
\newcommand{\dcb}{\delta_{\rm cav,b}}
\newcommand{\dcs}{\delta_{\rm cav,s}}
\newcommand{\hs}{h_{\rm s}}
\newcommand{\hb}{h_{\rm b}}
\newcommand{\hso}{h_{\rm s}^{\rm o}}
\newcommand{\hbo}{h_{\rm b}^{\rm o}}
\newcommand{\hsc}{h_{\rm s}^{\rm c}}
\newcommand{\hbc}{h_{\rm b}^{\rm c}}

\newcommand{\bso}{\beta_{\rm s}^{\rm o}}
\newcommand{\bbo}{\beta_{\rm b}^{\rm o}}
\newcommand{\bsc}{\beta_{\rm s}^{\rm c}}
\newcommand{\bbc}{\beta_{\rm b}^{\rm c}}
\newcommand{\md}{M_{\rm dust}}

\begin{document}
\title{The Structure of Pre-transitional Protoplanetary Disks I: Radiative Transfer Modeling of the Disk+Cavity in the PDS~70 System\footnote{Based on data collected at the Subaru Telescope, which is operated by the National Astronomical Observatory of Japan.}}

\shorttitle{Radiative transfer modeling of PDS~70}

\shortauthors{Dong et al.}

\author{Ruobing Dong\altaffilmark{1}, Jun Hashimoto\altaffilmark{2}, Roman Rafikov\altaffilmark{1}, Zhaohuan Zhu\altaffilmark{1}, Barbara Whitney\altaffilmark{3}, Tomoyuki Kudo\altaffilmark{4}, Takayuki Muto\altaffilmark{5}, Timothy Brandt\altaffilmark{1}, Melissa K. McClure\altaffilmark{6}, John Wisniewski\altaffilmark{7}, L. Abe\altaffilmark{8}, W. Brandner\altaffilmark{9}, J. Carson\altaffilmark{10}, S. Egner\altaffilmark{4}, M. Feldt\altaffilmark{9}, M. Goto\altaffilmark{9}, C. Grady\altaffilmark{11,12}, O. Guyon\altaffilmark{4}, Y. Hayano\altaffilmark{4}, M. Hayashi\altaffilmark{13}, S. Hayashi\altaffilmark{4}, T. Henning\altaffilmark{9}, K. W. Hodapp\altaffilmark{14}, M. Ishii\altaffilmark{4}, M. Iye\altaffilmark{2}, M. Janson\altaffilmark{1}, R. Kandori\altaffilmark{2}, G. R. Knapp\altaffilmark{1}, N. Kusakabe\altaffilmark{2}, M. Kuzuhara\altaffilmark{2}, J. Kwon\altaffilmark{2}, T. Matsuo\altaffilmark{15}, M. McElwain\altaffilmark{11}, S. Miyama\altaffilmark{16}, J.-I. Morino\altaffilmark{2}, A. Moro-Martin\altaffilmark{17}, T. Nishimura\altaffilmark{4}, T.-S. Pyo\altaffilmark{4}, E. Serabyn\altaffilmark{18}, H. Suto\altaffilmark{2}, R. Suzuki\altaffilmark{19}, M. Takami\altaffilmark{20}, N. Takato\altaffilmark{4}, H. Terada\altaffilmark{4}, C. Thalmann\altaffilmark{9}, D. Tomono\altaffilmark{4}, E. Turner\altaffilmark{1,21}, M. Watanabe\altaffilmark{22}, T. Yamada\altaffilmark{23}, H. Takami\altaffilmark{4}, T. Usuda\altaffilmark{4}, M. Tamura\altaffilmark{2}}

\altaffiltext{1}{Department of Astrophysical Sciences, Princeton University, Princeton, NJ 08544, rdong@astro.princeton.edu}
\altaffiltext{2}{National Astronomical Observatory of Japan, 2-21-1 Osawa, Mitaka, Tokyo 181-8588, Japan}
\altaffiltext{3}{Astronomy Department, University of Wisconsin-Madison, 475 N. Charter St., Madison, WI 53706, USA}
\altaffiltext{4}{Subaru Telescope, 650 North A'ohoku Place, Hilo, HI 96720, USA}
\altaffiltext{5}{Division of Liberal Arts, Kogakuin University, 1-24-2, Nishi-Shinjuku, Shinjuku-ku, Tokyo, 163-8677, Japan}
\altaffiltext{6}{Department of Astronomy, University of Michigan, 500 Church Street, Ann Arbor, MI 48105, USA}
\altaffiltext{7}{HL Dodge Department of Physics \& Astronomy, University of Oklahoma, 440 W Brooks St, Norman, OK 73019 USA}
\altaffiltext{8} {Laboratoire Lagrange, UMR7293, Universit\'e de Nice-Sophia Antipolis, 
CNRS, Observatoire de la C\^ote d'Azur, 06300 Nice, France}
\altaffiltext{9} {Max Planck Institute for Astronomy, Heidelberg, Germany}
\altaffiltext{10} {Department of Physics and Astronomy, College of
Charleston, 58 Coming St., Charleston, SC 29424, USA}
\altaffiltext{11}{ExoPlanets and Stellar Astrophysics Laboratory, Code 667, Goddard Space Flight Center, Greenbelt, MD 20771 USA}
\altaffiltext{12}{Eureka Scientific, 2452 Delmer, Suite 100, Oakland CA 96002, USA}
\altaffiltext{13} {Department of Astronomy, The University of Tokyo,
Hongo 7-3-1, Bunkyo-ku, Tokyo 113-0033, Japan}
\altaffiltext{14} {Institute for Astronomy, University of Hawaii, 640
North A'ohoku Place, Hilo, HI 96720, USA}
\altaffiltext{15} {Department of Astronomy, Kyoto University,
Kitashirakawa-Oiwake-cho, Sakyo-ku, Kyoto, 606-8502, Japan}
\altaffiltext{16} {Office of the President, Hiroshima University,
1-3-2 Kagamiyama, Higashi-Hiroshima, 739-8511, JAPAN}
\altaffiltext{17} {Departamento de Astrof\'isica, CAB (INTA-CSIC), 
Instituto Nacional de T\'ecnica Aeroespacial, Torrej\'on de Ardoz,
28850, Madrid, Spain}
\altaffiltext{18} {Jet Propulsion Laboratory, California Institute of Technology, Pasadena, CA, USA}
\altaffiltext{19} {TMT Observatory Corporation, 1111 South Arroyo
Parkway, Pasadena, CA 91105, USA}
\altaffiltext {20} {Institute of Astronomy and Astrophysics, Academia
Sinica, P.O. Box 23-141, Taipei 106, Taiwan}
\altaffiltext{21} {Kavli Institute for the Physics and Mathematics of the
Universe, The University of Tokyo, Kashiwa 227-8568, Japan}
\altaffiltext{22} {Department of Cosmosciences, Hokkaido University,
Sapporo 060-0810, Japan}
\altaffiltext{23} {Astronomical Institute, Tohoku University, Aoba,
Sendai 980-8578, Japan}

\clearpage

\begin{abstract}

Through detailed radiative transfer modeling, we present a disk+cavity model to simultaneously explain both the spectral energy distribution (SED) and Subaru $H$-band polarized light imaging for the pre-transitional protoplanetary disk PDS~70. In particular, we are able to match not only the radial dependence, but also the absolute scale, of the surface brightness of the scattered light. Our disk model has a cavity 65~AU in radius, which is heavily depleted of sub-micron-sized dust grains, and a small residual inner disk that produces a weak but still optically thick Near-IR excess in the SED. To explain the contrast of the cavity's edge in the Subaru image, a factor of $\sim$1000 depletion for the sub-micron-sized dust inside the cavity is required. The total dust mass of the disk may be on the order of $10^{-4}M_\odot$, only weakly constrained due to the lack of long wavelength observations and the uncertainties in the dust model. The scale height of the sub-micron-sized dust is $\sim6$ AU at the cavity edge, and the cavity wall is optically thick in the vertical direction at $H$-band. PDS~70 is not a member of the class of (pre-)transitional disks identified by Dong et al., whose members only show evidence of the cavity in the millimeter-sized dust but not the sub-micron-sized dust in resolved images. The two classes of (pre-)transitional disks may form through different mechanisms, or they may simply be at different evolution stages in the disk clearing process.

\end{abstract}

\keywords{protoplanetary disks --- stars: pre-main sequence --- stars: individual (PDS 70) --- radiative transfer --- circumstellar matter}


\section{Introduction}\label{sec:intro}

Recent years have witnessed a rising interest in a special kind of protoplanetary disks, in which a giant central cavity is present and reveals itself in the spectral energy distribution \citep[SED, e.g.][]{str89,skr90,cal05,fur06,esp07a,esp07b,esp08}), or in sub-mm interferometry observations \citep[e.g.][]{pie06,hug07,bro09,hug09,ise10,and11,mat12,cie12b,ise12,and12}. Depending on whether a small residual optically thick inner disk is left at the center, these objects could be classified into two categories --- pre-transitional disks (with an optically thick residual inner disk, \citealt{esp10}), and transitional disks (without an optically thick residual inner disk, \citealt{esp07b}). Studying these objects at multiple wavelengths is of great interests, because the inner disk clearing process may be signposts of planet formation and disk evolution in general \citep{zhu11,wil11,dod11,bir12,cie12a}.

So far, the study of (pre-)transitional disks was done primarily via measurements of the SED and resolved images at sub-mm wavelengths. More recently, direct imaging at optical to near infrared (NIR) started playing a crucial role. Protoplanetary disks usually have dust grains with sizes ranging from sub-micron (called small dust in this study), similar to the pristine interstellar medium (ISM) dust, to mm or larger (big dust), forming as a result of grain growth and coagulation \citep{dul05,bir12}. The SED of a disk is degenerate with many disk parameters, since it is spatially integrated over the entire disk. Resolved images at sub-mm, on the other hand, provide more detailed information, but they are only sensitive to the distribution of big dust due to its dominance in opacity at these wavelengths. Using 8-m class ground based telescopes equipped with Adaptive Optics and/or coronagraph system, direct imaging at optical to NIR wavelengths provides resolved disk maps with high spatial resolution ($\sim0.06\arcsec$) and small inner working angle ($\lesssim0.2\arcsec$). These maps are directly related to the distribution of small dust.

An ongoing survey scale project, the Strategic Explorations of Exoplanets and Disks with Subaru (SEEDS, \citealt{tam09}), is directly imaging a large sample of protoplanetary disks at NIR ($J$, $H$, and $K$) bands in a systematic way, using the High-Contrast Coronographic Imager for Adaptive Optics (HiCIAO, \citealt{suz10}). SEEDS disk observations specialize in taking polarized intensity (PI=$\sqrt{Q^2+U^2}$, which $Q$ and $U$ are components in the Stokes vector) images of disks, which greatly enhances our ability to probe disk structure (especially at the inner part), by utilizing the fact that the central source is usually not polarized, so that the stellar residual in PI images is much smaller than in full intensity (FI) images \citep{per04,hin09,qua11}. The signal in NIR imaging usually comes from scattering of starlight off the surface of the disk, since dust at separations of $\sim10$ AU in the disk is not hot enough to contribute significantly at NIR wavelengths. A number of protoplanetary disks have been studied in SEEDS (LkCa 15, \citealt{tha10}, AB Aur, \citealt{has11}, SAO 206462, \citealt{mut12}, MWC 480, \citealt{kus12}, and UX Tau A, \citealt{tan12}), and more are on the way.

Recently, \citet{dong12} pointed out that in a large sample of (pre-)transitional disks whose central cavity have been confirmed in resolved images at sub-mm, SEEDS did not find the cavities in their NIR images. This discrepancy has been interpreted as evidence for the cavity existence {\it only in big dust}, with the small dust having a {\it continuous} distribution in surface density from the outer disk all the way to the inner working angle of SEEDS ($\sim0.\!\!''1-0.\!\!''15$, or about $\sim15-20$ AU in Taurus). This may be explained by the dust filtration model proposed by \citet{paa06,ric06,zhu12}. In their model, the pressure bump in the disk acts like a filter, which filters through the small dust but traps the big dust, resulting in the depletion of big dust inside the cavity.

In this article, we study the structure of the pre-transitional disk PDS~70 using detailed radiative transfer modeling, following the observational paper by \citet{has12}. PDS~70 is a K5 type weak-lined T Tauri star located at the Centaurus star-forming region (coordinate 14 08 10.15 -41 23 52.5), with an age $<10$Myr and a distance estimated to be $\sim140$pc \citep{gre02,ria06,met04,has12}. The photometry of this star has been measured at multiple wavelengths (e.g. \citealt{met04,kes05}), and its disk has been detected in scattered light by VLT $K$-band imaging \citep{ria06}. SEEDS directly imaged PDS~70 in $H$-band using the polarization differential imaging (PDI) mode on Feb. 27, 2012. The Subaru observation and the data reduction were described in detail in \citet[which also presented the Gemini $L^\prime$-band imaging]{has12}. Unlike most previous objects of this type, the cavity in PDS~70 is {\it first} detected by SEEDS NIR imaging. And unlike the objects in \citet{dong12}, it indeed has a cavity in the small dust, which also contradicts the predictions in various dust filtration models mentioned above.

Radiative transfer modeling of (pre-)transitional disks has been mostly used to fit the SED and sub-mm observations in the past (e.g. \citealt{hug09,and11,mat12,and12}). Here we demonstrate the power of modeling in NIR, by producing synthetic disk images in scattered light and comparing them with observations. Sensitive to dust at different sizes and locations, SED, sub-mm observations, and NIR imaging can be used to probe the disk structure in different and complementary ways. Particularly, NIR imaging can provide effective constraints on the scale height of the disk, and the depletion factor for the small dust inside the cavity, neither of which is well constrained using the other two types of observations \citep{and11,dong12}.

This paper is the first in a series, in which we study the structures of (pre-)transitional disks by modeling SEEDS NIR imaging data and observations at other wavelengths. The paper is arranged in the following way. In Section~\ref{sec:modeling} we introduce the method for our radiative transfer modeling. The fiducial model of PDS~70 is presented in Section~\ref{sec:fiducial}, where we simultaneously fit both its SED and scattered light image. The constraints on various disk+cavity parameters from current observations are discussed in Section~\ref{sec:constraint}, where we focus on the ones which are directly probed by NIR imaging and are important in revealing the disk evolution. We discuss the possible formation mechanisms of (pre-)transitional disks in Section~\ref{sec:discussion}, followed by a short summary in Section~\ref{sec:summary}.



\section{Radiative transfer modeling}\label{sec:modeling}

We use the Monte Carlo radiative transfer (MCRT) code developed by B. Whitney (2012, in prep. see also \citealt{whi03a,whi03b,rob06}) to model the SED and SEEDS $H$-band PI imaging of PDS 70. The SED data collected from literature, the SEEDS observation, and its data reduction are described in detail in \citet{has12}. MCRT simulations are run with $4\times10^7$ photons for high S/N images. Disk setup is largely adopted from the model in \citet{dong12}, which is briefly summarized here. We construct an axisymmetric disk 200~AU in radius on a $600\times200$ grid in spherical coordinates ($R,\theta$), where $R$ is the radial component and $\theta$ is the poloidal component ($\theta=0^\circ$ is the disk mid-plane). We uniformly deplete the central region of the disk to form a cavity with radius $\rcav$, i.e. reduce dust surface density by a constant factor within this radius. The disk is assumed to be at $\sim$140~pc. The inner radius of the disk (within the cavity) is self-consistently determined by dust sublimation ($T_{\rm sub}\sim1600$ K). The accretion rate of PDS~70 is considered to be very low, since the object has been identified as a weak line T Tauri object with an H$\alpha$ equivalent width of 2\AA~\citep{gre02}. For the central source, we assume a pre-main sequence star of spectral type K5, radius 1.35 $R_\odot$, mass 1.0 $M_\odot$, and temperature 4500~K, as suggested by \citet{gre02}. The surface density $\Sigma(R)$ in the outer disk is taken as
\begin{equation}
\Sigma(R)=\Sigma_0\frac{R_c}{R}e^{-R/R_c},\ {\rm at\ }R\geq\rcav,
\label{eq:sigma}
\end{equation}
where $R_c$ is a characteristic scaling length assumed to be 50 AU, and the normalization $\Sigma_0$ is determined by the total dust mass of the disk $\md$. The choices of the surface density radial profile and $R_c$ are common in protoplanetary disk modeling \citep{wil11,and11}, and represent a snapshot of a solution for a fully viscous disk with a constant $\alpha$ parameter (as in \citealt{sha73} $\alpha$ disk model) and $T \propto R^{-1/2}$ \citep{har98}. Inside the cavity, the disk is uniformly depleted, with surface density going as
\begin{equation}
\Sigma(R)=\dc\Sigma_0\frac{R_c}{R}e^{-R/R_c},\ {\rm at\ }R<\rcav,
\label{eq:sigma}
\end{equation}
where $\dc$ is the constant depletion factor (which may be different for different dust populations).

Various grain evolution models predict that the pristine dust grains in the disk coagulate, grow to bigger size, and subsequently settle to the disk mid-plane \citep{dul04a,dul04b,dul05}. To take this effect into account, we assume a two component model of dust distribution: a thick disk with small grains (sub-micron-sized), and a thin disk with large grains (up to $\sim$mm-sized). Below we will use subscripts ``b'' and ``s'' for various quantities relating to big and small dust, respectively, and superscripts ``o'' and ``c'' to indicate the outer disk and the cavity, respectively (for example, $\Sigma_s^c$ represents the surface density of the small dust inside the cavity). Both big and small grains are assumed to have a Gaussian density profile in the vertical direction,
\begin{equation}
\rho(R,z)=\frac{\Sigma(R)}{\sqrt{2\pi}h} e^{-z^2/2h^2},
\label{eq:rhorz}
\end{equation}
where $h$ is the scale height, with $\hb$ (scale height of the big dust) assumed to be much smaller than $\hs$ (scale height of the small dust). We note that the scale heights are provided to the code as input parameters, instead of being self-consistently determined from the disk temperature calculated in the simulations. In our fiducial model shown below, we check this assumption and find that the input is consistent with the output (Section~\ref{sec:fiducial}). Radially, both scale heights vary with radius as
\begin{equation}
h\propto R^\beta,
\label{eq:b}
\end{equation}
where $\beta$ is the constant power law index. Our experiments show that as long as the big dust is settled to the disk mid-plane (i.e. $\hb\ll\hs$), the details of its vertical distribution hardly affect the details of NIR images and SED, while both crucially depend on the spatial distribution of the small dust, as we will show below in Section~\ref{sec:scaleheight}.

The total mass of the big dust in the disk is $f\times\md$, where $f$ is a variable parameter. For the small dust, we assume a size distribution as in the standard ISM dust model from \citet{kim94} up to a maximum size of 0.2~$\mu$m (roughly a power-law size distribution $n(s)\propto s^p$ with power index $p\sim3.5$). The dust composition is assumed to be $70\%$ mass in silicate and $30\%$ mass in graphite, with the properties for both adopted from \citet{lao93}. Preferentially forward scattering (i.e. Mie scattering) is assumed. For the big dust we primarily use Model 2 from \citet{woo02} (from now on called steep-big-dust), which assumes a power-law size distribution $n(s)\propto s^p$ with $p=3.5$ up to a maximum size of 1~mm, and a composition of amorphous carbon and astronomical silicates. In Section~\ref{sec:mass} Model 3 from \citet{woo02} is also tried (from now on called flat-big-dust), whose only difference from the steep-big-dust model is that it assumes $p=3.0$ (``flatter'' size distribution, so more mass at the large size end). As a consequence, the flat-big-dust model has lower opacity than the steep-big-dust model at wavelengths $\lambda\lesssim1$mm. The opacity of these dust models are shown in Figure~\ref{fig:kappa}.

PDS~70 shows a cavity with radius $\sim60-70$AU, and the disk is inclined by about $45^\circ-50^\circ$ \citep{has12}. In our fiducial model shown below we choose 65 AU ($\sim0.\!\!''46$) and $45^\circ$ for the two, as they produce model SED and image which match observations reasonably well (Section~\ref{sec:fiducial}). We note that a mild deviation from these ``fiducial'' values can be tolerated without much difficulty (i.e. $\pm5$ AU in cavity size and $\pm5^\circ$ in geometry). These parameters are fixed below to keep the models simple.

We produce $H$-band polarized images from MCRT simulations. To obtain realistic images which can be directly compared with SEEDS observations, the raw simulated images are post-processed, as described in detail in \citet{dong12}. The SEEDS observations of PDS~70 were conducted in PDI mode without a coronagraph \citep{has12}, so that we produce mock images in the same mode in this work. The raw model images of the entire disk+star is convolved by a real $H$-band SEEDS point spread function (PSF). The inner working angle of the images, $\psi_{\rm in}$, is assumed to be $0.\!\!''15$ in radius ($\sim$21 AU at the estimated distance of PDS~70). We measure the surface brightness radial profile (SBRP) of the disk along major and minor axes, by calculating the average SB of the pixels within $\pm22.5^\circ$ on each side of the axes at various radial bins $0.\!\!''05$ in width, the same as we measured for the SBRP of the SEEDS PDS~70 image \citep{has12}.

\section{The fiducial disk model for PDS~70}\label{sec:fiducial}

Here we present a fiducial disk model for PDS~70 to fit all the observations. The disk parameters in this model are listed in Table~\ref{tab:model}, and the surface density of the small dust is plotted in Figure~\ref{fig:sigma}. In summary, the disk has a heavily depleted cavity whose radius is 65 AU $\rcav$, with a depletion factor of 1000 for both the big and small dust ($\dcb$ and $\dcs$, we note that only $\dcs$ is constrained by the current observations, not $\dcb$, see below). The inner-most disk (on AU scales) and the cavity wall are both optically thick in the vertical direction at 1~$\mu$m, representing the peak of the stellar spectra. The vertical optical depth due to the small dust is $\sim14$ at the inner edge of the disk ($\sim0.06$ AU), and $\sim4$ at the cavity wall. On the other hand, the gap in between the inner rim and the outer disk (i.e. from $\sim$1 AU to $\rcav$) is optically thin to stellar radiation, which justifies the classification of PDS~70 as a pre-transitional disk \citep{esp10}. The total dust mass $\md$ is $3\times10^{-5}M_\odot$, and most of it is in the big dust, with a big-to-small-dust ratio at about 30:1 (i.e. $f=0.97$). However, as we will discuss in Section~\ref{sec:mass}, the constraints on the big dust are rather weak (including the total amount, its structure, and the big-to-small-dust ratio), given the current data. We assume scale height power law index 1.2 for both dust populations ($\beta_{\rm b}$ and $\beta_{\rm s}$), both inside and outside the cavity (indicated by subscripts ``c'' and ``o''), reasonably close to the canonical value $\sim1.25-1.3$ for irradiated disks \citep{chi97,har98}. The gas in the disk is in hydro-static equilibrium in the vertical direction, so that $h_{\rm gas}$ is set by the vertical temperature profile, through $h_{\rm gas}\approx\cs/\Omega$ and $\cs\approx\sqrt{k_BT/\mu}$, where $\cs$ is the sound speed, $\Omega$ is the rotational angular velocity, $T$ is the mid-plane temperature, and $\mu$ is the average molecular weight. Various grain settling models predict that the small grains tend to be well coupled with the gas \citep{dul04b}, as a result the two would share similar vertical distribution, $\hs\sim h_{\rm gas}$. The small dust scale height is 6.0 AU at $\rcav$, both inside the cavity ($\hsc$) and outside the cavity ($\hso$), which corresponds to a mid-plane temperature of 28.2K. This value is close to the output from the radiative transfer calculation, 31.8K. The big dust scale height inside and outside the cavity  ($\hbc$ and $\hbo$) is 1/5 of the value for the small dust.

The model $H$~band PI images are shown in Figure~\ref{fig:fiducial-image}, along with the observed SEEDS image. The SED and SBRP (defined in Section~\ref{sec:modeling}) of the image are shown in Figure~\ref{fig:fiducial-sed-sbrp}. Our model SED matches the observations very well. For the scattered light image, our fiducial model looks very similar to SEEDS image, both revealing a clear cavity with size $\sim0.\!\!''5$ on the major axes. If fit by an ellipse, the disk center has an offset $\sim9$ AU from the star, roughly along the minor axis, and towards the far side of the disk, which is due to the back illumination of the wall. This offset is consistent with observation (the measured offset in SEEDS image is $\sim6$AU in the same direction, see \citealt{has12}). The small (bright) structures that appear close to the inner working angle in the SEEDS image is probably artificial, mostly likely caused by observational noise or the stellar residual in polarized light. The right side of the Subaru image is slightly brighter (in scattered light) than its left counterpart, which is not reproduced by our axisymmetric fiducial model. This asymmetry may be intrinsic, caused by differences in grain properties, or disk structure (i.e. scale height or the amount of dust), on the two sides. We note that since we focus on studying the global scale structure, particularly the general properties of the cavity (i.e. its size and depletion factor), using axisymmetric models, we do not address the local details and non-axisymmetric structures in this work.

Quantitatively, we reach good agreement with the measurement of the SBRP along the major axis. The only obvious deviation happens at large distance ($\gtrsim\!\!''1$, well beyond the cavity edge), where unlike the model, the observed image flattens out to a (roughly constant) background noise. Due to the axisymmetry of our model, the SBRPs along both directions at its semi-major axis are the same, while observationally these profiles are slightly different, due to reasons discussed above. Here we emphasize that we achieve good agreement not only for the radial dependence (i.e. the slope), but also for the {\it absolute scale} of the surface brightness (the vertical axis in all the SBRP plots in this study is in actual physical units, and the curves were not rescaled).

The agreement between our model and SEEDS observation on the minor axes is not perfect, but nevertheless the two agree on major qualitative features: the surface brightness on the far side of the disk peaks at a larger radius, and it decreases outward slower than on the near side. The flux in polarized scattered light is determined by the product of polarization fraction, and the intensity of the full intensity (i.e. PI=(PI/FI)$\times$FI). Along the major axis, the scattering angle is nearly 90$^\circ$, which results in a maximum polarization fraction (PI/FI) due to the phase function of small dust. On the other hand, along the minor axis the scattering happens closer to the center, and at angles off 90$^\circ$, resulting in small PI/FI. On the far side, both the upper and lower edge of the cavity are visible, so that the ring is wider; on the near side the lower edge is blocked by the outer disk, however forward scattering leads to a bigger FI there. We note that generally speaking, the comparison along the major axis is more valuable in constraining the disk+cavity structure, because disk is spatially more extended in this direction so that it is better resolved by observations with a fixed spatial resolution in all directions. Also, along the minor axis the features of interests (i.e. cavity edge) are present at closer separation from the center (also $\psi_{\rm in}$), where the photon noise is generally larger.

We note that the parameters and geometry of our fiducial model come from an overall consideration of fitting the SED, NIR image, and radial profile of the scattered light simultaneously, instead of simply measuring from the image as in \citet{has12}. Also, the fiducial model presented here is by no means unique, i.e. the only one which can provide a good fit to all the observations, as the constraints on some of the disk+gap parameters are rather weak (Section~\ref{sec:constraint}).

At last, we comment on the previous VLT $K$-band imaging of PDS~70 reported by \citet{ria06}, and their derived disk model. Observationally, the inclination and position angle found by \citet{ria06} is similar to our results \citep{has12}. However, our Subaru $H$-band PI images provide far more details and are at a higher quality than the VLT $K$-band full intensity images presented by \citet{ria06}, which detected the disk in scattered light, but did not reveal a clear giant cavity structure, and showed a jet structure that is not present in our observations. \citet{ria06} employed a smooth disk model with radius larger than 500 AU, and a total dust mass between 0.001 to 0.002 $M_\odot$ to reproduce the observations, which is one order of magnitude lager than in our fiducial model. The difference is mostly caused by the fact that the inner region is heavily depleted in our models but not in \citet{ria06}.

\section{Constraints on various disk and cavity parameters}\label{sec:constraint}

The fiducial model presented above provides a reasonably good fit to observations, and gives us a basic idea about the disk+cavity structure. In this section, we intend to determine the constraints on some of the parameters from modeling and fitting, so that we can understand what these observations are really telling us. In general, disk properties at short wavelengths (i.e. scattered light images and NIR excess in the SED) are sensitive to the spatial distribution of the small dust, while this dependence shifts to the big dust at long wavelengths (i.e. FIR excess and sub-mm observations), due to the difference in opacity of the two populations (Figure~\ref{fig:kappa}). Since there are too many free parameters in our model, it is not realistic to vary every one of them and study their effects. Rather, we narrow our scope to few {\it key} parameters:
\begin{enumerate}
\item The ones which are crucial in revealing the evolution of protoplanetary disk, and the formation mechanisms of (pre-)transitional disks. These are the depletion factor of both dust populations, and the total mass of the disk.
\item The ones which can {\it only} be effectively constrained by scattered light images, such as the scale height and depletion factor of the small dust, which has a large degeneracy in SED and sub-mm observations. Since SEEDS virtually opens a new window to systematically study a large uniform sample of disks using scattered light images, we intend to provide an example to demonstrate the power of NIR imaging for probing disk structure.
\end{enumerate}

Below we study the role played by these factors both in disk SED and scattered light imaging, by exploring the corresponding parameter space around the fiducial model. We examine the cavity depletion factors for both dust populations ($dcb$ and $dcs$) in Section~\ref{sec:depletion}, the total dust mass ($\md$) in Section~\ref{sec:mass}, and the scale height of the small dust in the cavity and the outer disk ($\hsc$, $\bsc$, $\hso$, and $\bso$) in Section~\ref{sec:scaleheight}. Models and their parameters in each section are listed and described in the corresponding block in Table~\ref{tab:model}. For the SED we focus on comparing with the photometry data, and for the scattered light we only look at the SBRP along the major axis.

\subsection{The depletion factor inside the cavity}\label{sec:depletion}

The cavity depletion factor of the small dust $\dcs$ could be constrained by both the SED and the scattered light image. Qualitatively, depleting the cavity more in small dust reduces the short wavelength excess on the SED, and enhances the contrast of the cavity in the image. Technically, the SED is more sensitive to the depletion in the inner part of the cavity (at AU scales) since most of the short wavelength excess is produced there, while the scattered light images are more sensitive to the depletion in the outer part of the cavity, where the contrast of the cavity edge is produced (the inner part of the disk within $\psi_{\rm in}$ cannot be directly accessed in imaging observations). As a result, the two could in principle be constrained ``independently''. However, to simplify our discussion, we use a uniform instead of radius-dependent $\dcs$ (and $\dcb$), and note that this treatment does not affect our conclusion (see \citealt{dong12} for a discussion on radius-dependent $\dc$).

Figure~\ref{fig:depletion}a shows the effect of $\dcs$ on the SED ($\dcb$ is locked to $\dcs$, but this hardly affects our result, as shown below). The surface density profile of the small dust ($\Sigma_{\rm s}$) for these models is plotted in Figure~\ref{fig:sigma}. Whether $2-20\mu$m excess strongly correlates with $\dcs$ sensitively depends on whether the innermost disk (at AU scale) is optically thick or not (characterized at the peak of the stellar spectra, $\sim1\mu$m). When the innermost disk is optically thick (pre-transitional disks), the IR excess is almost independent of the amount of small dust inside the cavity (the fiducial model and model SCM1). However, once $\Sigma_{\rm s}$ at the inner disk decreases below the optically thick limit, and enters the transitional disk phase, the $2-20\mu$m excess drops significantly as a result of the decreasing $\dcs$ (model SCM2 to SCM4). If the inner disk is completely depleted (SCM4), excess below $\sim$10 $\mu$m disappears, and a nearly blackbody thermal component peaking at $\sim40\mu$m clearly reveals itself, which arises from the cavity wall. Though not as isolated as in SCM4, this wall emission signal is prominent in all models. Comparing with observation, we conclude that the NIR excess in PDS~70 is consistent with an optically thick innermost disk, which requires $\dcs>10^{-4}$.

Figure~\ref{fig:depletion}b shows the SBRP of the convolved image for the above models. As expected, decreasing $\dcs$ makes the inner disk fainter. Moreover, the blocking effect due to the inner disk is reduced, so that a bigger area of the cavity wall is illuminated, and more starlight (rays closer to the disk mid-plane) reaches the wall and outer disk, increasing their brightness in scattered light. Similarly, increasing $\dcs$ tends to wipe out the signal of the cavity in scattered light (i.e. the bump at $\sim0.\!\!''05$), and makes the SBRP smoother. The observed SEEDS SBRP is broadly consistent with $\dcs\sim10^{-3}$, while a modest deviation (i.e. a factor of $\sim3$) around this value could be tolerated without too much difficulty.

On the other hand, the cavity depletion factor of the big dust could only be directly constrained by sub-mm properties of the disk \citep{and11,cie12b,ise12,mat12}. Figures~\ref{fig:depletion}c,d show that changing $\dcb$ hardly affects either the disk SED or NIR image (we fix the small dust to isolate the effect, i.e. the depletion factor for the small dust component inside of the cavity is fixed to $\dcs\sim10^{-3}$ in all the models plotted here). We leave the constraint on $\dcb$ to future sub-mm observations.

To conclude, $\dcs\sim10^{-3}$ is needed for the models to be consistent with both the SED and SEEDS image. The constraint is modest, roughly half a dec around the fiducial value.

\subsection{The total dust mass of the disk}\label{sec:mass}

The total dust mass of the disk $\md$ is a quantity crucial for a thorough understanding of disk evolution. In general, $\md$ is determined by the big dust since it dominates in mass over the small dust, and for (pre-)transitional disks like PDS~70, most of the mass resides in the outer disk since the inner disk is heavily depleted. $\md$ is best constrained by sub-mm or mm observations, because the disk is usually optically thin at these wavelengths, so $\md$ can be calculated from the measured sub-mm or mm flux, assuming a dust temperature and opacity model (i.e. Equation 2 and 3 in \citealt{wil11}). However, the total disk mass derived in this way normally contains large uncertainties, introduced by poorly unconstrained dust opacity and gas-to-dust-ratio, if converting dust mass to total gas mass \citep{pan08}.

For PDS~70, we constrain the disk mass using the longest wavelength photometric data point available, which is at 160$\mu$m. Our models are all vertically optically thin (sometimes only marginally) at 160$\mu$m given our dust models. Models BOM1 and BOM2 (both have the steep-big-dust model) in Figure~\ref{fig:mass}a show the SED dependence on $\md$. While the fiducial choice of $\md=3\times10^{-5}M_\odot$ agrees well with the observed photometry at 160$\mu$m, a factor of $\sim3$ deviation in $\md$ from this value leads to a factor $\sim2$ difference in 160$\mu$m flux, while the SED at wavelengths shorter than $\sim100\mu$m is largely unchanged.

One issue which deserves special attention in this exercise of $\md$ determination is the effect of the big dust model. Since the long wavelength flux in the optically thin regime is proportional to the opacity of the dust at that wavelength, changing the dust model for the big dust has a direct impact on $\md$. Figure~\ref{fig:mass} shows two examples in which the flat-big-dust model is assumed instead of the steep-big-dust model. As mentioned in Section~\ref{sec:modeling} and shown in Figure~\ref{fig:kappa}, the flat-big-dust has a flatter size distribution, resulting in smaller opacity at wavelengths shorter than $\sim$1mm. Model BOM3-flat has an identical set of disk parameters as the fiducial model. While the difference in SED between the two models is negligible at $\lambda\lesssim100\mu$m, the IR excess in BOM3-flat sharply drops below the fiducial model at longer wavelengths. To pull the 160$\mu$m flux back to the observed value, the disk mass needs to rise to $1.5\times10^{-4}M_\odot$, as in model BOM4-flat. On the other hand, since all the BOM models have identical spatial distribution for the small dust, their scattered light images are almost the same, as shown in Figure~\ref{fig:mass}b.

We conclude that the total dust mass of PDS~70 $\md$ is probably on the order of $10^{-4}M_\odot$, depending on the dust model for the big dust: a flatter grain size distribution for the big dust corresponds to a larger $\md$. Observations at longer wavelengths are needed to distinguish different big dust models and to pin down the disk mass.

\subsection{The scale height of the small dust}\label{sec:scaleheight}

The scale height of the small dust $\hs$, which dominates the absorption of starlight, is a central quantity involved in determining almost all the observable properties of the disk\footnote{The big dust is generally considered to have settled to the disk mid-plane, and its opacity at the peak of the stellar spectrum is much lower than that of the small dust. As a result, the detailed vertical distribution of the big dust does not have a prominent effect on the SED and the scattered light image.}. Scale height and total mass of the small dust determine the shape of the disk surface, on which the stellar radiation is absorbed and re-emitted at longer wavelengths. Part of that emission escapes and gets observed, and part of it goes deeper into the disk interior to heat the disk, subsequently being reprocessed into disk emission at even longer wavelengths. Despite its importance, unfortunately very few types of observations can determine $\hs$ in a straightforward manner (see the example in \citealt{mut12}, where the shape of the spiral waves in SAO 206462 is used to estimate $\hs$). However, scattered light is an excellent tool for probing $\hs$, since its signal comes right from the disk scattering surface, which is directly determined by $\hs$.

In this section, we use several models to investigate the effects of $\hs$ on disk SED and NIR image. Figure~\ref{fig:scaleheight}a,b show the result of models SOS1-4, in which the scale height at the outer disk $\hso$ is varied. We find that the radial dependence of $\hso$ has a minimal effect on both the SED and scattered light image (models SOS1 and SOS2), while the overall scale of $\hso$ plays a prominent role in both the SED and NIR image. A much larger area of the cavity wall in model SOS4, whose $\hso$ is 1.5 times the fiducial value, is directly exposed to the stellar radiation, due to its larger height. As a result, its SED is a factor of $\sim2$ higher at the peak of the wall emission ($\sim40\mu$m) compared with the fiducial model and the observed SED, and the cavity edge is $\sim50\%$ brighter.

Figures~\ref{fig:scaleheight}c,d show the effect of $\hs$ inside the cavity ($\hsc$, model SCS1-4). Similar to $\hso$, the absolute scale of the scale height plays a bigger role than its radial dependence. Model SCS1 and SCS3, both having a thinner innermost disk, produce less NIR excess and slightly less scattered light from the inner disk. Also, since less starlight is blocked by the innermost disk and more of it reaches the cavity edge, the two have more MIR excess around 40$\mu$m and a slightly brighter cavity edge at $H$-band. On the contrary, model SCS4 has a thicker disk inside the cavity, resulting in more short wavelength excess and less long wavelength excess, due to less starlight reaching the outer disk. Its inner disk is also brighter and outer disk is fainter in the scattered light image. In general, the effect of $\hsc$ on the SED is modest, echoing the finding in Section~\ref{sec:depletion}, that once the inner disk becomes optically thick, the NIR excess is insensitive to the distribution of small dust.

In sum, the scale height of the small dust at $\rcav$, $h/R\sim0.09$, is relatively well determined by both the SED and the scattered light image (note that this value is consistent with the output temperature from radiative transfer calculation, Section~\ref{sec:fiducial}). On the other hand, the constraint on the radial dependence of $\hs$ is weaker, as a broad range of $\beta$ from 1.15 to 1.25 (both inside and outside the cavity) in the parameter space we explored here does not contradict the observations.



\section{Indication on the formation of (pre-)transitional disks}\label{sec:discussion}

PDS~70 is a special pre-transitional disk, in the sense that unlike most of its previous cousins, whose cavities were usually first inferred from the shape of the SED and then confirmed by resolved sub-mm images, the cavity in PDS~70 was {\it first} found in SEEDS scattered light image (see also \citealt{hon10} for AB Aur). Follow up interferometer observations at long wavelengths, for example using the Submillimeter Array (SMA) or the Atacama Large Millimeter Array (ALMA), are needed to check the existence of the cavity in the big dust. If it is indeed confirmed, then this object will represent a class of (pre-)transitional disks which are in clear contrast to the ones discussed in \citet{dong12}. \citet{dong12} reported the discovery of a class of (pre-)transitional disks, whose cavities are confirmed using the SMA interferometer but not seen in SEEDS scattered light images, despite the fact that the inner working angle of the SEEDS images is small enough to reveal these cavities (if they exist). In addition, observations at long wavelengths will help to determine the total dust mass of the disk, which is currently poorly contained to be $\sim10^{-4}M_\odot$ due to the lack of long wavelength observations and the uncertainties in dust model.

There are only a handful of objects whose cavities have been revealed in resolved images at multiple wavelengths covering a broad range. A special group of them are circumbinary disks (with the secondary being a stellar or sub-stellar object), such as GG Tau. The cavity in GG Tau has been found in the optical \citep{kri05}, NIR \citep{ito02}, mm \citep{pie11}, and CO line emission \citep{gui99}, all at roughly the same position. The formation mechanism of cavities in circumbinary systems has been studied by \citet{art94}. It is thought that the gravitational interaction between the massive (sub-)stellar secondary and the disk naturally truncates the disk, and the cavity formed in this way exists in all disk components, including gas and dust at different sizes. The imaging at $L^\prime$ band shown in \citet{has12} has put an upper limit $\sim30-50M_J$ for any possible companion at the radii of interest (assuming their age of the system and the (sub-)stellar object evolution model), which rules out the possibility of a stellar mass companion. Future observations with better sensitivity and contrast performance are needed to answer whether a sub-stellar companion exists (i.e. a brown dwarf). On the other hand, the possibility of gap opening due to multiple Jovian planets has been explored by \citet{zhu11,dod11}. The general picture in that scenario, that a wide deep gap is opened while a small optically thick inner-most disk is left at the center, qualitatively agrees with the observations of PDS~70, though it is generally harder to directly image Jovian planet(s) in a bright protoplanetary disk.

The dust filtration model has been proposed by \citet{paa06} and \citet{ric06}, in which the pressure maximum at the planet-induced gap outer edge acts like a filter, so that big grains are trapped but small grains penetrate into the inner disk. As a result, it is predicted that the cavity must be depleted of big dust grains, but with a significant amount of small dust particles left inside. PDS~70, with its heavily depleted cavity in small dust grains, is inconsistent with the dust filtration model alone. On the other hand, \citet{zhu12} proposed a dust filtration + dust growth model as a possible explanation for the transitional disk GM Aur. The grain growth and coagulation at the innermost disk may turn the small grains into big grains, resulting in a (pre-)transitional-disk-like NIR excess (see also \citealt{bir12}). The dust and gas components of the disk are decoupled in their model and provide an explanation for both the moderate accretion rate of GM Aur and its strong near-IR deficit. Furthermore, this dust size dependent filtration model may explain the different gap properties between near-IR and sub-mm reported in \citet{dong12}. However, unlike GM Aur, PDS~70 is a WTTS, suggesting a very low gas accretion rate. Together with the clear cavity revealed in the NIR imaging, a consistent picture emerges for PDS~70, that both the small dust and the gas are heavily depleted at the inner disk, and there is no clear evidence for decoupling between the two. On the other hand, without sub-mm/mm observations, we do not know if small and big dust are decoupled or not.

Nevertheless, following the dust filtration + dust growth model, we can still provide an explanation for the PDS~70 class of objects. They may start as systems modeled in \citet{dong12}, which do have a significant amount of small dust inside the cavity and the NIR cavity is not present. Later on, grain growth and coagulation, which happen at a faster rate in the inner disk due to high density and short dynamical timescale, gradually spread over the entire inner disk, eventually leading to the formation of a small dust cavity. If this scenario is true, then these two kinds of (pre-)transitional disks are just at different evolutionary stages in their cavity clearing process --- the ones in \citet{dong12} are at an early stage, while PDS~70 is at a later stage. Consequently, we should expect to see a statistical ``time delay'' between the two types of objects. This time delay is partly supported by the facts that PDS~70 is a WTTS, and that it is a relatively old system. This argument can be further tested if a large sample of objects in both categories with accurate age determination could be provided by future observations. Also, another possible strong evidence in favor of this scenario would be the detection of objects in the intermediate phase of this process, such as (1) objects with a ``partially cleared'' small dust cavity whose edge is between the center and the edge of the big dust cavity, or (2) objects with a radius-dependent cavity depletion factor for the small dust (probably smaller at inner disk), showing that the clearing process is moving outward. Furthermore, the byproduct of grain growth and coagulation --- a slight enhancement of the big dust signal inside the cavity at later stage --- may be observable using a sub-mm interferometer with very high sensitivity, such as ALMA \citep{dong12}.



\section{Summary}\label{sec:summary}

We carry out a study of the disk+cavity structure for PDS~70, using radiative transfer modeling to fit both the observed SED and the SEEDS polarized scattered light image at $H$-band of this object. Good agreement with observations is achieved in our models. The disk has a giant cavity at its center with radius 65 AU. The small dust (sub-micron-sized) inside the cavity is depleted by a factor of $\sim1000$, resulting in a low density but still vertically optically thick innermost disk, producing a pre-transitional-disk-like NIR excess. This heavy depletion is also needed to explain the surface density depression inferred from the scattered light image. The scale height of the small dust at the cavity edge is $\sim$6 AU, constrained by both the SED and the image, which is consistent with the output mid-plane temperature of the disk. On the other hand, the total mass of the disk can be estimated only crudely to be on the order of $10^{-4}M_\odot$, due to the lack of sub-mm and mm data, and the degeneracy of the dust models for the big dust in the SED. This quantity, along with the cavity depletion factor for the big (mm-sized) dust, would be determined by future observations at longer wavelengths.

Unlike most previously classified (pre-)transitional disk, the cavity in PDS~70 is identified in NIR scattered light, which only informs us of the depletion in the small dust but tells us little about the spatial distribution of the big dust. Pending the confirmation of the cavity in the big dust by radio interferometer observations, PDS~70 may be a prototype of its group, in which the cavity is seen in both dust populations. It is in clear contrast with the (pre-)transitional disks discussed in \citet{dong12}, where the small dust cavities were not seen outside the inner working angle of the scattered light images. The (pre-)transitional disks with or without a NIR cavity may be formed through different mechanisms (i.e. binary vs planets or grain growth), or they may just be caught at different evolutionary stage in their disk clearing process. Observational predictions for both mechanisms are made in Section~\ref{sec:discussion}, and more objects with multi-wavelengths observations in both categories are needed to reveal the nature of (pre-)transitional disks.


\section*{Acknowledgments}

R.D. thanks Edwin Bergin, Tilman Birnstiel, Brendan Bowler, Nuria Calvet, L. Ilsedore Cleeves, Bruce Draine, Catherine Espaillat, Lee Hartmann, Shu-ichiro Inutsuka, Andrea Isella, James Owen, Thomas P. Robitaille, Fredrik Windmark, and Yanqin Wu for useful discussions. The authors are also grateful to the anonymous referee who helped improve the manuscript. This work is partially supported by NSF grant AST 0908269 (R. D., Z. Z., and R. R.), AST 1008440 (C. G.), AST 1009314 (J. W.), AST 1009203 (J. C.), NASA grant NNX22SK53G (L. H.), Sloan Fellowship (R. R.), and World Premier International Research Center Initiative (WPI Initiative), MEXT, Japan (E. L. T.). Part of this research was carried out at the Jet Propulsion Laboratory, California Institute of Technology, under a contract with the National Aeronautics and Space Administration.


\clearpage

\input{tab.tex}

\input{fig.tex}

\end{document}

%% file: tab.tex
\begin{landscape}
\begin{deluxetable}{ccccccccccccc}
\tabletypesize{\scriptsize}
\tablewidth{0pc}
\tablecaption{Disk Models}
\tablehead{
\colhead{Name} &
\colhead{$\md$} &
\colhead{$f$} &
\colhead{$\hbo$(100AU)} &
\colhead{$\bbo$} &
\colhead{$\hbc$(100AU)} &
\colhead{$\bbc$} &
\colhead{$\dcb$} &
\colhead{$\hso$(100AU)} &
\colhead{$\bso$} &
\colhead{$\hsc$(100AU)} &
\colhead{$\bsc$} &
\colhead{$\dcs$}\\
\colhead{(1)} &
\colhead{(2)} &
\colhead{(3)} &
\colhead{(4)} &
\colhead{(5)} &
\colhead{(6)} &
\colhead{(7)} &
\colhead{(8)} &
\colhead{(9)} &
\colhead{(10)} &
\colhead{(11)} &
\colhead{(12)} &
\colhead{(13)}
}

\startdata
Fiducial & $3\times10^{-5}$ & 0.967 & 2 & 1.2 & 2 & 1.2 & $10^{-3}$ & 10 & 1.2 & 10 & 1.2 & $10^{-3}$\\
\hline
\multicolumn{13}{c}{The Depletion Factor Inside the Cavity}\\
SCM1 & $3\times10^{-5}$ & 0.967 & 2 & 1.2 & 2 & 1.2 & $10^{-2}$ & 10 & 1.2 & 10 & 1.2 & $10^{-2}$\\
SCM2 & $3\times10^{-5}$ & 0.967 & 2 & 1.2 & 2 & 1.2 & $10^{-4}$ & 10 & 1.2 & 10 & 1.2 & $10^{-4}$\\
SCM3 & $3\times10^{-5}$ & 0.967 & 2 & 1.2 & 2 & 1.2 & $10^{-5}$ & 10 & 1.2 & 10 & 1.2 & $10^{-5}$\\
SCM4 & $3\times10^{-5}$ & 0.967 & 2 & 1.2 & 2 & 1.2 & 0 & 10 & 1.2 & 10 & 1.2 & 0\\
BCM1 & $3\times10^{-5}$ & 0.967 & 2 & 1.2 & 2 & 1.2 & $10^{-2}$ & 10 & 1.2 & 10 & 1.2 & $10^{-3}$\\
BCM2 & $3\times10^{-5}$ & 0.967 & 2 & 1.2 & 2 & 1.2 & 0 & 10 & 1.2 & 10 & 1.2 & $10^{-3}$\\
\hline
\multicolumn{13}{c}{The Total Dust Mass of the Disk}\\
BOM1 & $10^{-4}$ & 0.99 & 2 & 1.2 & 2 & 1.2 & $2.9\times10^{-4}$ & 10 & 1.2 & 10 & 1.2 & $10^{-3}$\\
BOM2 & $10^{-5}$ & 0.9 & 2 & 1.2 & 2 & 1.2 & $3.3\times10^{-2}$ & 10 & 1.2 & 10 & 1.2 & $10^{-3}$\\
BOM3-flat & $3\times10^{-5}$ & 0.967 & 2 & 1.2 & 2 & 1.2 & $10^{-3}$ & 10 & 1.2 & 10 & 1.2 & $10^{-3}$\\
BOM4-flat & $1.5\times10^{-4}$ & 0.993 & 2 & 1.2 & 2 & 1.2 & $2.0\times10^{-4}$ & 10 & 1.2 & 10 & 1.2 & $10^{-3}$\\
\hline
\multicolumn{13}{c}{The Scale Height of the Small dust}\\
SOS1 & $3\times10^{-5}$ & 0.967 & 2 & 1.2 & 2 & 1.2 & $10^{-3}$ & 10 & 1.25 & 10 & 1.2 & $10^{-3}$\\
SOS2 & $3\times10^{-5}$ & 0.967 & 2 & 1.2 & 2 & 1.2 & $10^{-3}$ & 10 & 1.15 & 10 & 1.2 & $10^{-3}$\\
SOS3 & $3\times10^{-5}$ & 0.967 & 2 & 1.2 & 2 & 1.2 & $10^{-3}$ & 7.5 & 1.2 & 10 & 1.2 & $10^{-3}$\\
SOS4 & $3\times10^{-5}$ & 0.967 & 2 & 1.2 & 2 & 1.2 & $10^{-3}$ & 15 & 1.2 & 10 & 1.2 & $10^{-3}$\\
\\
SCS1 & $3\times10^{-5}$ & 0.967 & 2 & 1.2 & 2 & 1.2 & $10^{-3}$ & 10 & 1.2 & 10 & 1.25 & $10^{-3}$\\
SCS2 & $3\times10^{-5}$ & 0.967 & 2 & 1.2 & 2 & 1.2 & $10^{-3}$ & 10 & 1.2 & 10 & 1.15 & $10^{-3}$\\
SCS3 & $3\times10^{-5}$ & 0.967 & 2 & 1.2 & 2 & 1.2 & $10^{-3}$ & 10 & 1.2 & 7.5 & 1.2 & $10^{-3}$\\
SCS4 & $3\times10^{-5}$ & 0.967 & 2 & 1.2 & 2 & 1.2 & $10^{-3}$ & 10 & 1.2 & 15 & 1.2 & $10^{-3}$\\
\enddata
\tablecomments{Column (1): Name of the models. The first, second, and third letter indicates whether the model is about (1) big dust (B) or small dust (S); (2) outer disk (O) or cavity (C); and (3) mass (M) or scale height (S). In block ``The Total Mass of the Disk'', parameters are chosen in such a way that all the disk properties are the same except the mass of big dust in the outer disk (which essentially determines the total disk mass), and ``flat'' indicates that the flat-big-dust model is used instead of the steep-big-dust model (BOM3-flat has an identical set of parameters as the fiducial model). Column (2): Total dust mass of the disk, in unit of $M_\odot$. Column (3): Mass fraction of big dust in total dust. Columns (4), (6), (9), and (11): Scale height $h$ at 100 AU, in unit of AU. Subscripts ``b'' and ``s'' indicate big and small dust, respectively. Superscripts ``o'' and ``c'' indicate outer disk and cavity, respectively. Columns (5), (7), (10), and (12): Power index $\beta$ in Equation~\ref{eq:b} in various disk components. Column (8) and (13): Depletion factor of the big and small dust disk.}
\label{tab:model}
\end{deluxetable}
\end{landscape}

%% file: fig.tex
\begin{figure}[tb]
\begin{center}
\epsscale{0.45} \plotone{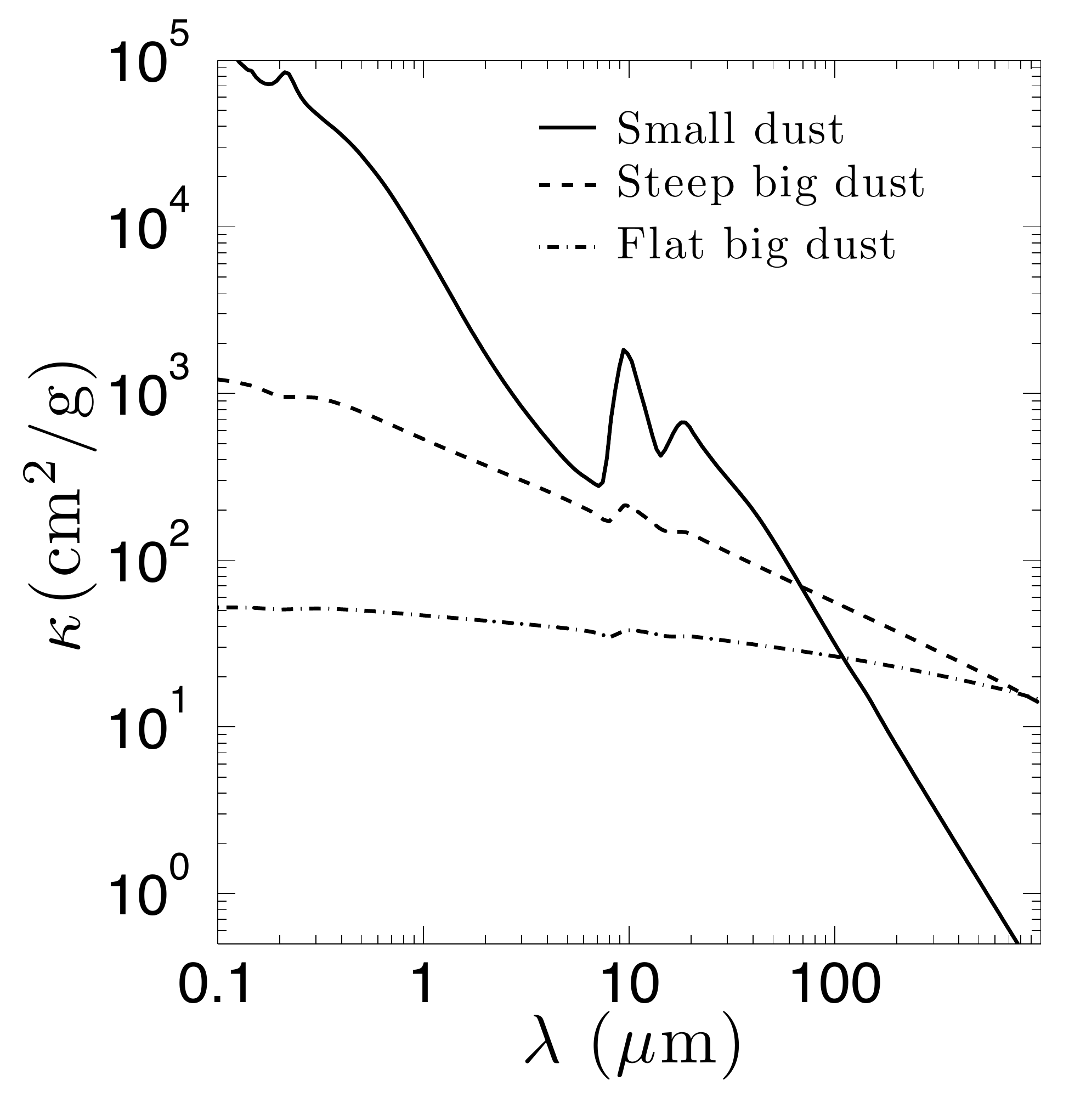}
\end{center}
\figcaption{Opacity $\kappa$ for the dust models used in this study (showing the dust opacity only).
\label{fig:kappa}}
\end{figure}

\begin{figure}[tb]
\begin{center}
\epsscale{0.45} \plotone{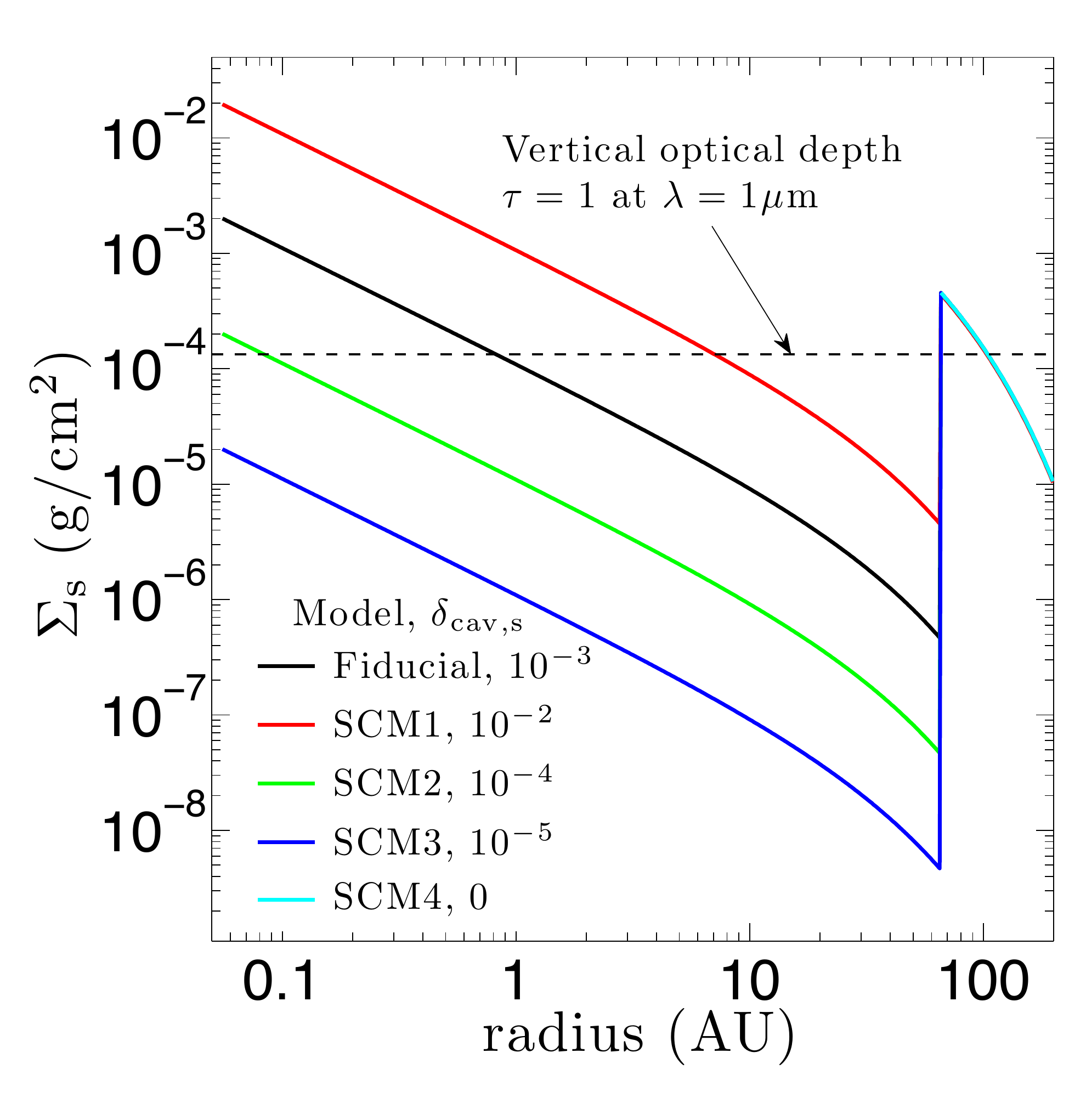}
\end{center}
\figcaption{Surface density radial profile of the small dust $\Sigma_{\rm s}$, for models shown in Figure~\ref{fig:depletion}a,b. The horizontal dashed line indicates where the vertical optical depth is 1 at 1$\mu$m (characteristic wavelength of the stellar radiation). The fiducial model and model SCM1 have vertically optically thick innermost disks, and models SCM3 and SCM4 have optically thin innermost disks, while model SCM2 is on the margin.
\label{fig:sigma}}
\end{figure}

\begin{figure}[tb]
\begin{center}
\epsscale{0.4} \plotone{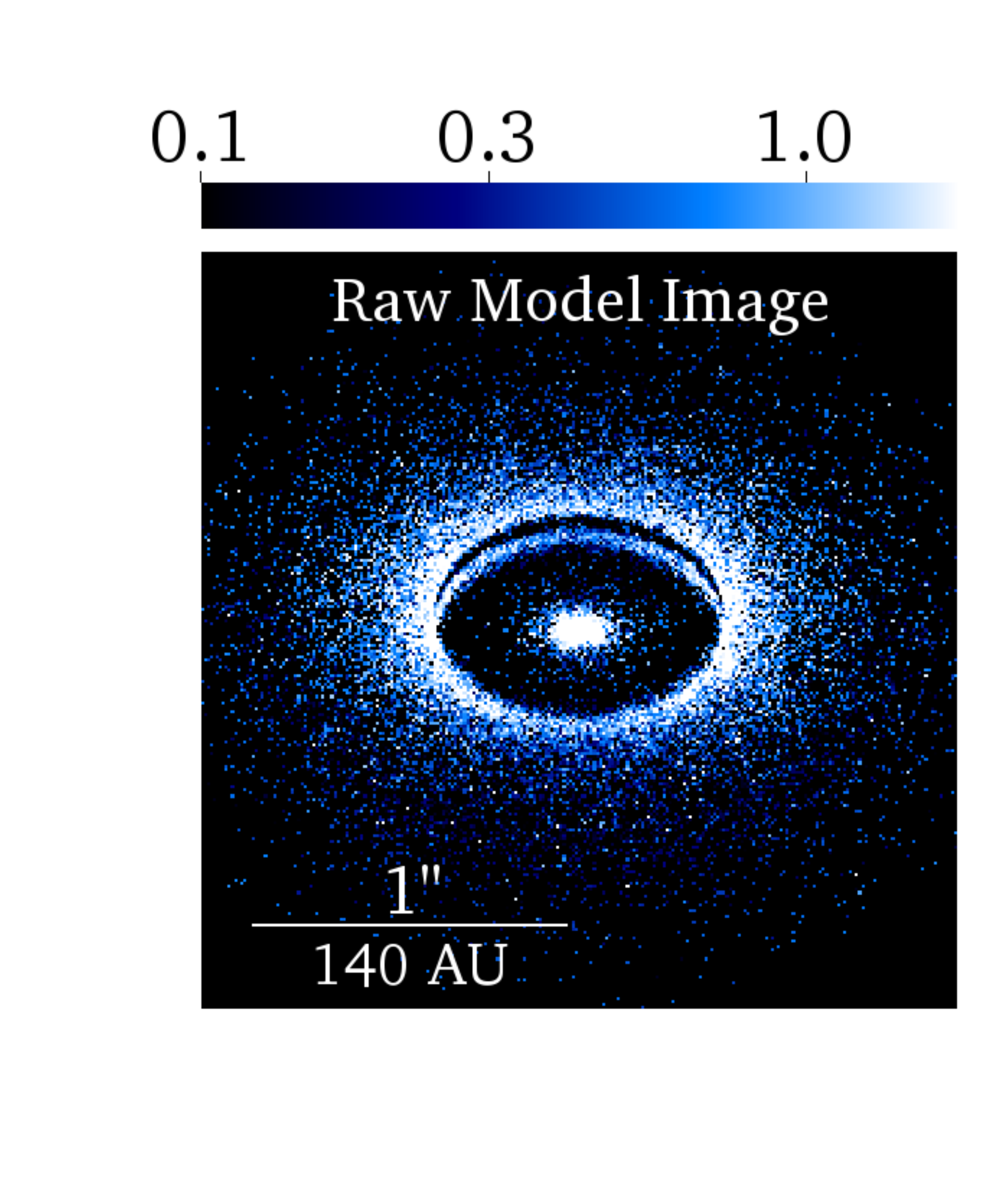} \plotone{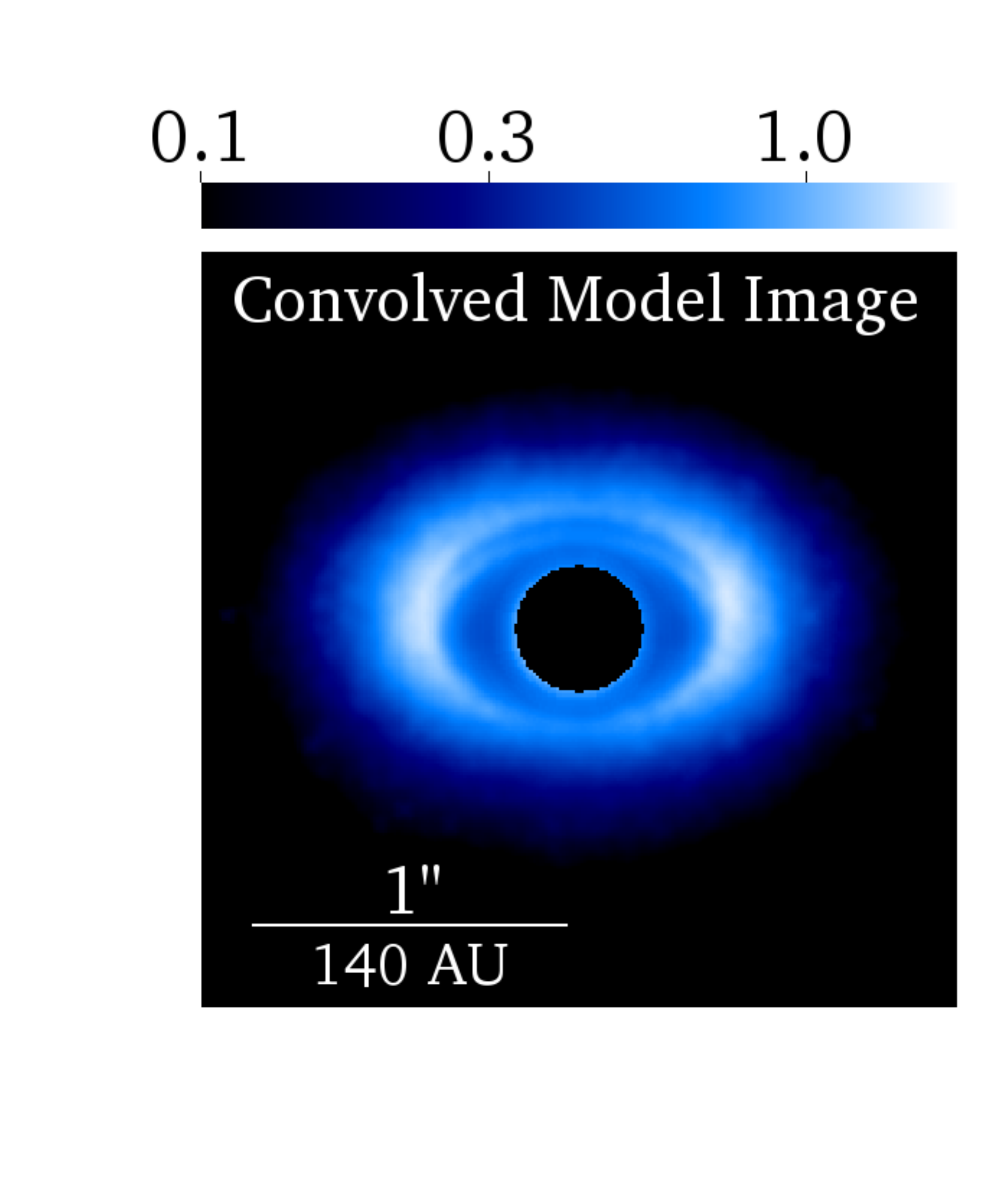} \plotone{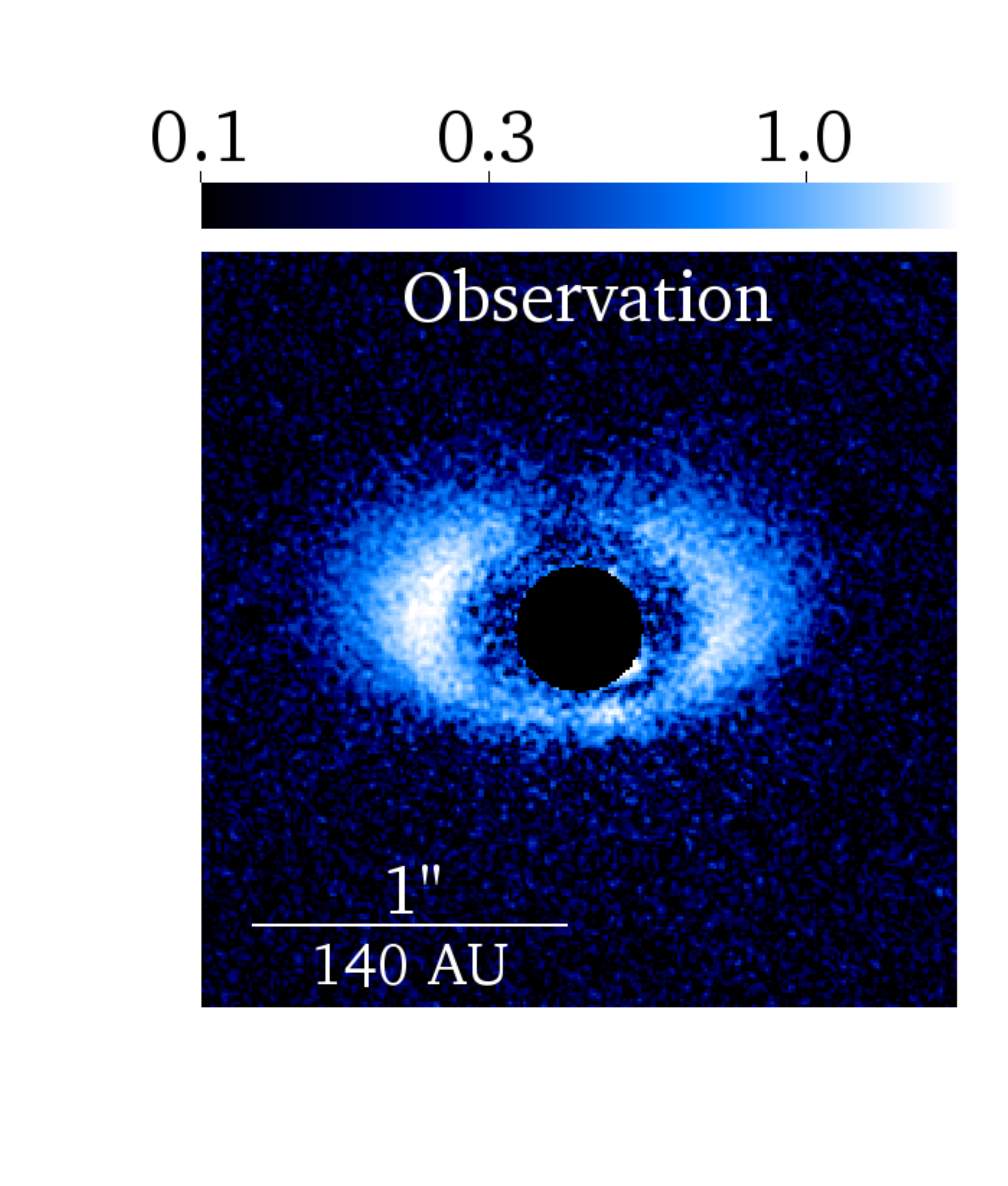} \plotone{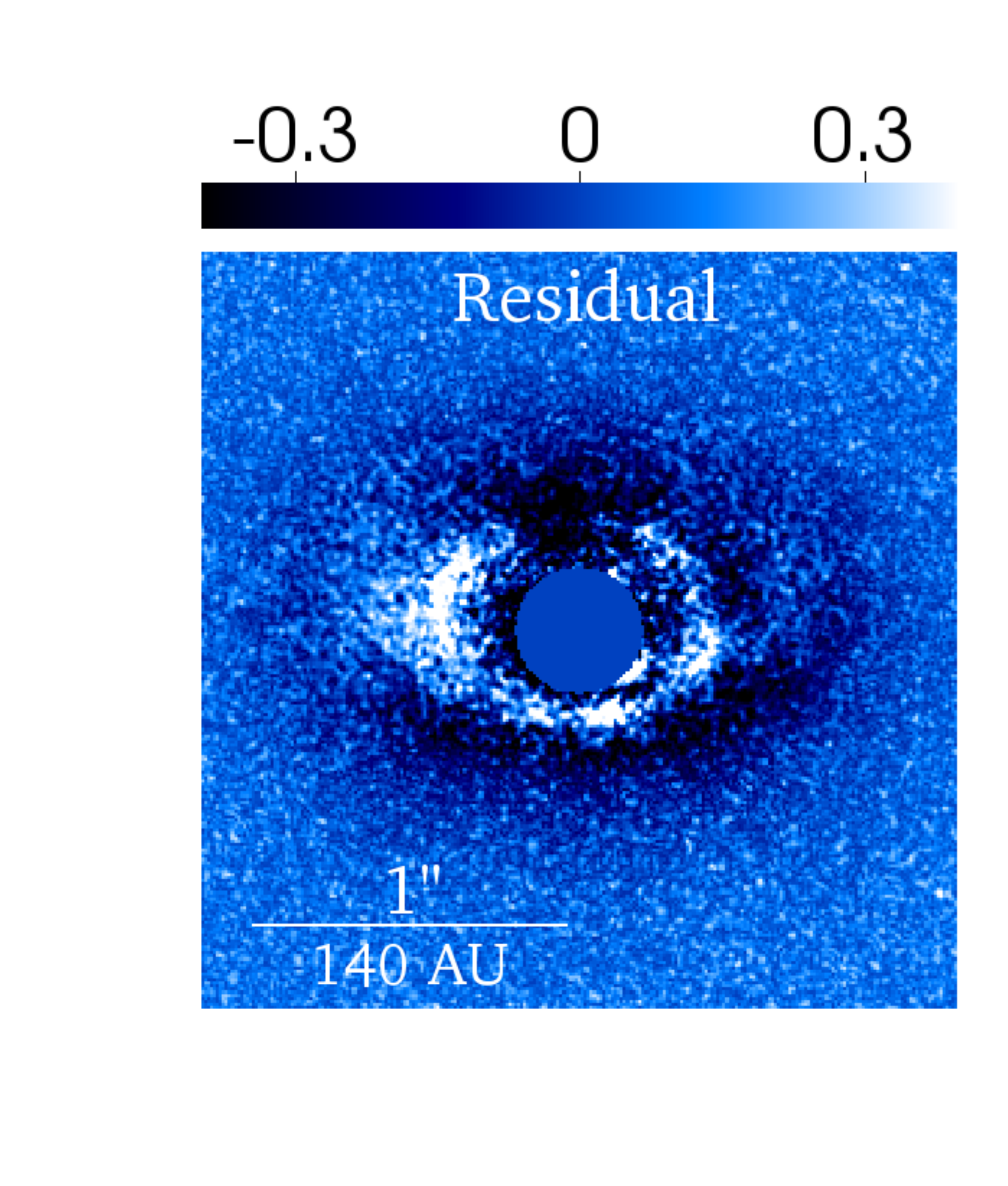}
\end{center}
\figcaption{Comparison of the polarized scattered light images at $H$-band between our fiducial model and SEEDS observation of PDS 70. Images show the surface brightness. The convolved model image is the raw model image convolved by a HiCIAO PSF (Section~\ref{sec:modeling}). The observed SEEDS image is not smoothed. The residual image is the convolved model image subtracted from the SEEDS image. All images are oriented in such a way that the far side of the disk ($21^\circ$ from west to north) is up. The mask at the center in the convolved model image and SEEDS image indicate a $0.\!\!''2$ (radius) inner working angle. Labels are in unit of mJy/arcsec$^2$ (the residual image is on linear scale, while the other three are on log scale with the same color scheme). Our fiducial model matches the large scale characteristics in the observation well, although some local and asymmetric features are not reproduced perfectly. See Section~\ref{sec:fiducial} for details.
\label{fig:fiducial-image}}
\end{figure}

\begin{figure}[tb]
\begin{center}
\epsscale{0.45} \plotone{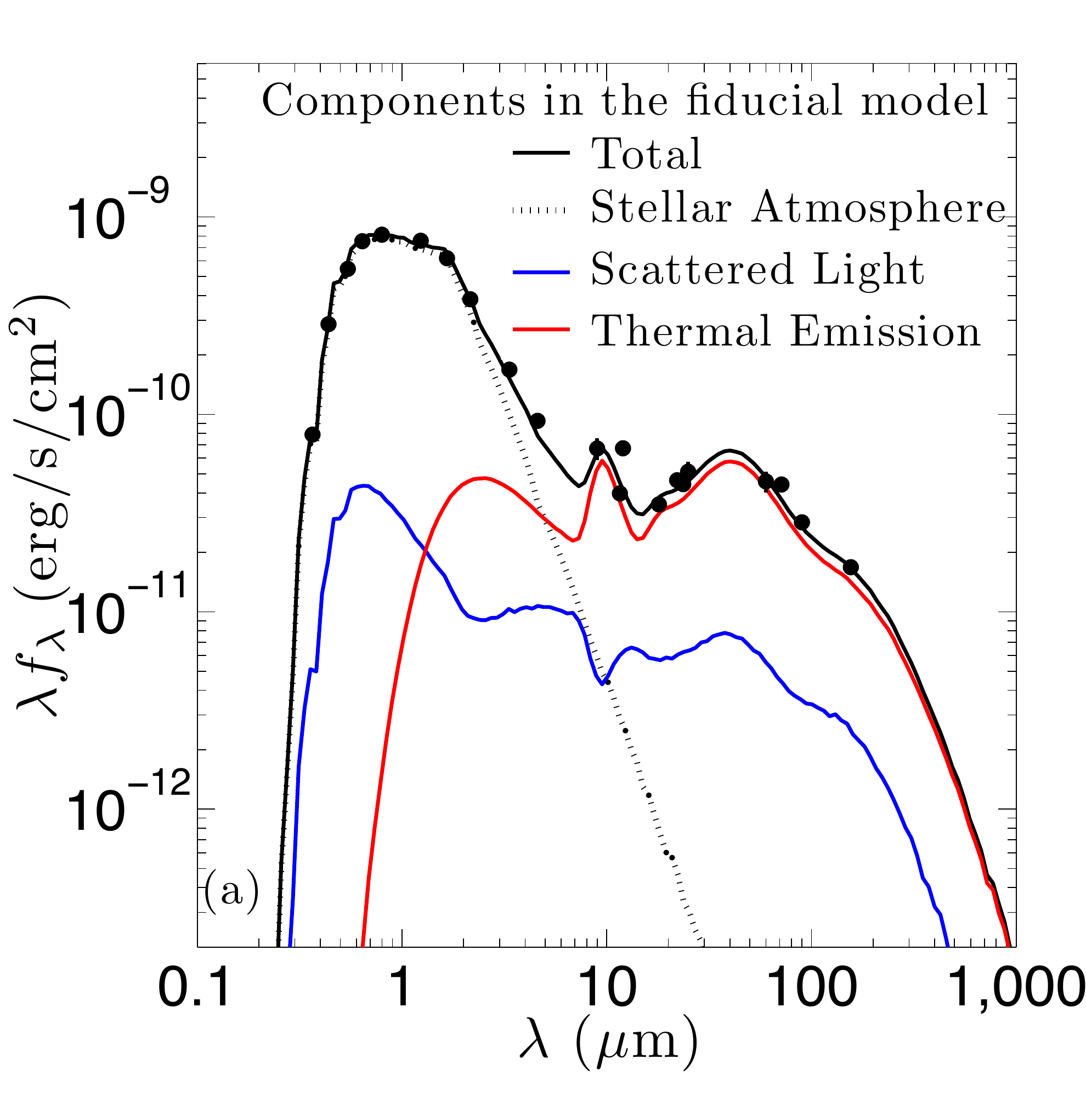} \plotone{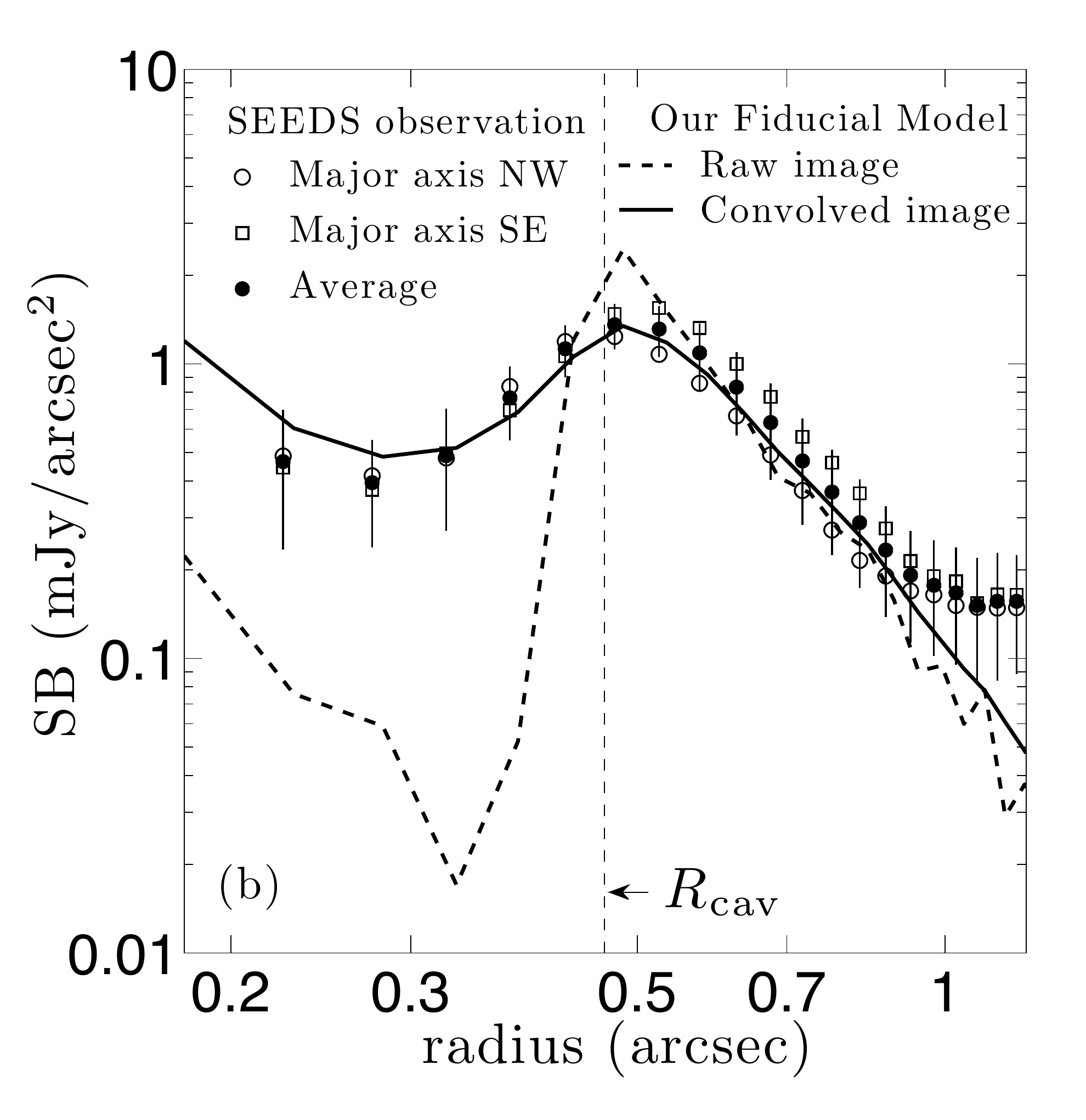} \plotone{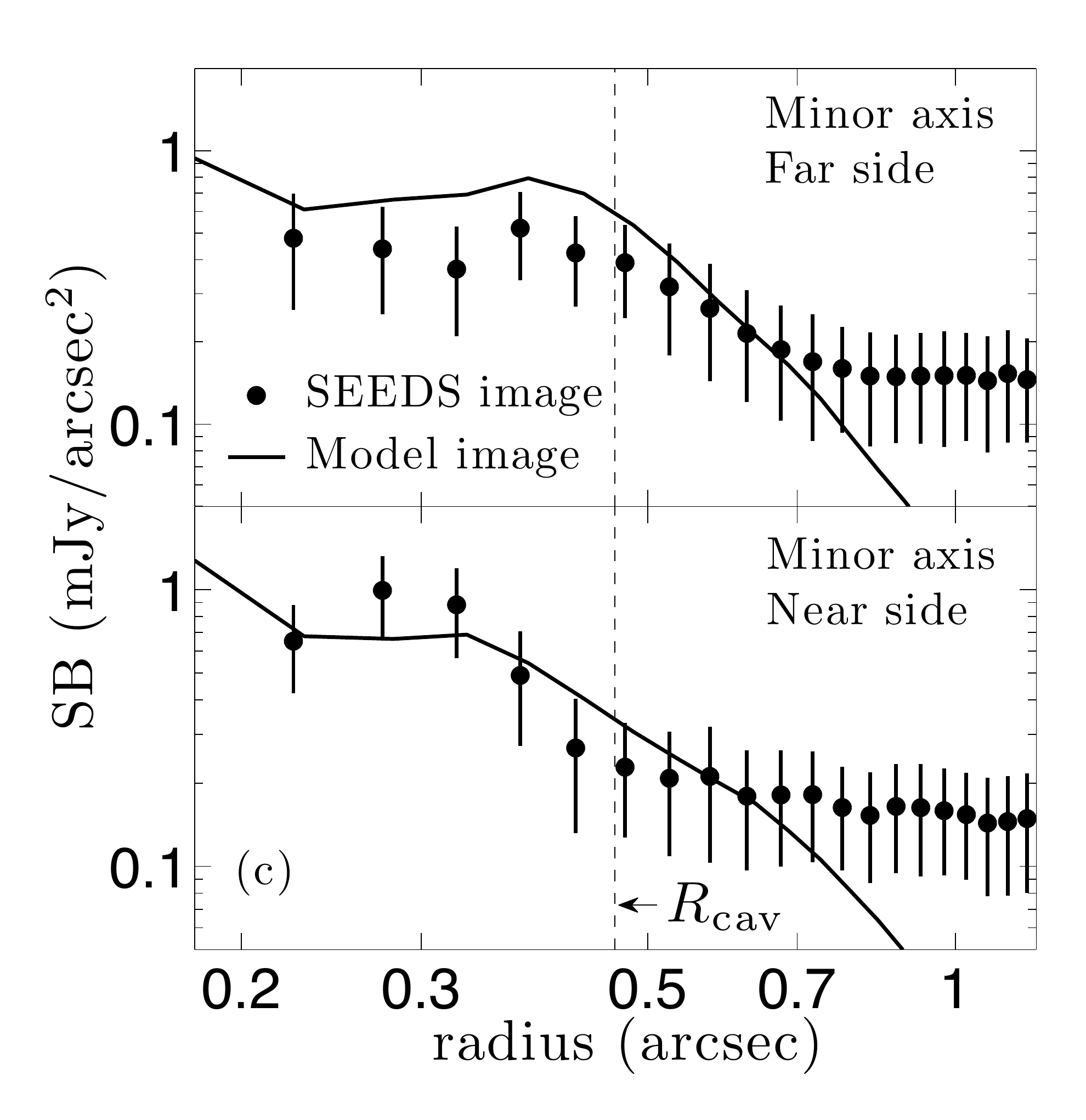}
\end{center}
\figcaption{Comparison between observations and our fiducial model on the SED (a), SBRP of the scattered light image along major axes (b), and SBRP along minor axes (c). Observational data is from paper I. In panel (a), the black dots are photometry data points (the vertical small bar indicates the error), and different components in the model SED are labeled. In panel (b) and (c), the black dots are the measured SEEDS SBRP, with error bars overplotted as vertical ticks. The individual observed SBRPs along the two directions of the semi-major axis are slightly different, and both are plotted as well. The fiducial model agrees well with the observations (particularly on the absolute scale of the surface brightness in the SEEDS image). In panel (c), due to the difficulty in measuring SBRP along the minor axes (Section~\ref{sec:fiducial}), the agreement is worse than in panel (b), but nevertheless the basic characteristic trend is still well matched, that the SBRP on the far side of the disk peaks at larger radii and decreases slower than on the near side, due to the back illumination. Model SBRP in (b) and (c) do not flatten out to a (constant) background noise at large radii, as in the observed SBRP.
\label{fig:fiducial-sed-sbrp}}
\end{figure}

\begin{figure}[tb]
\begin{center}
\epsscale{0.45} \plotone{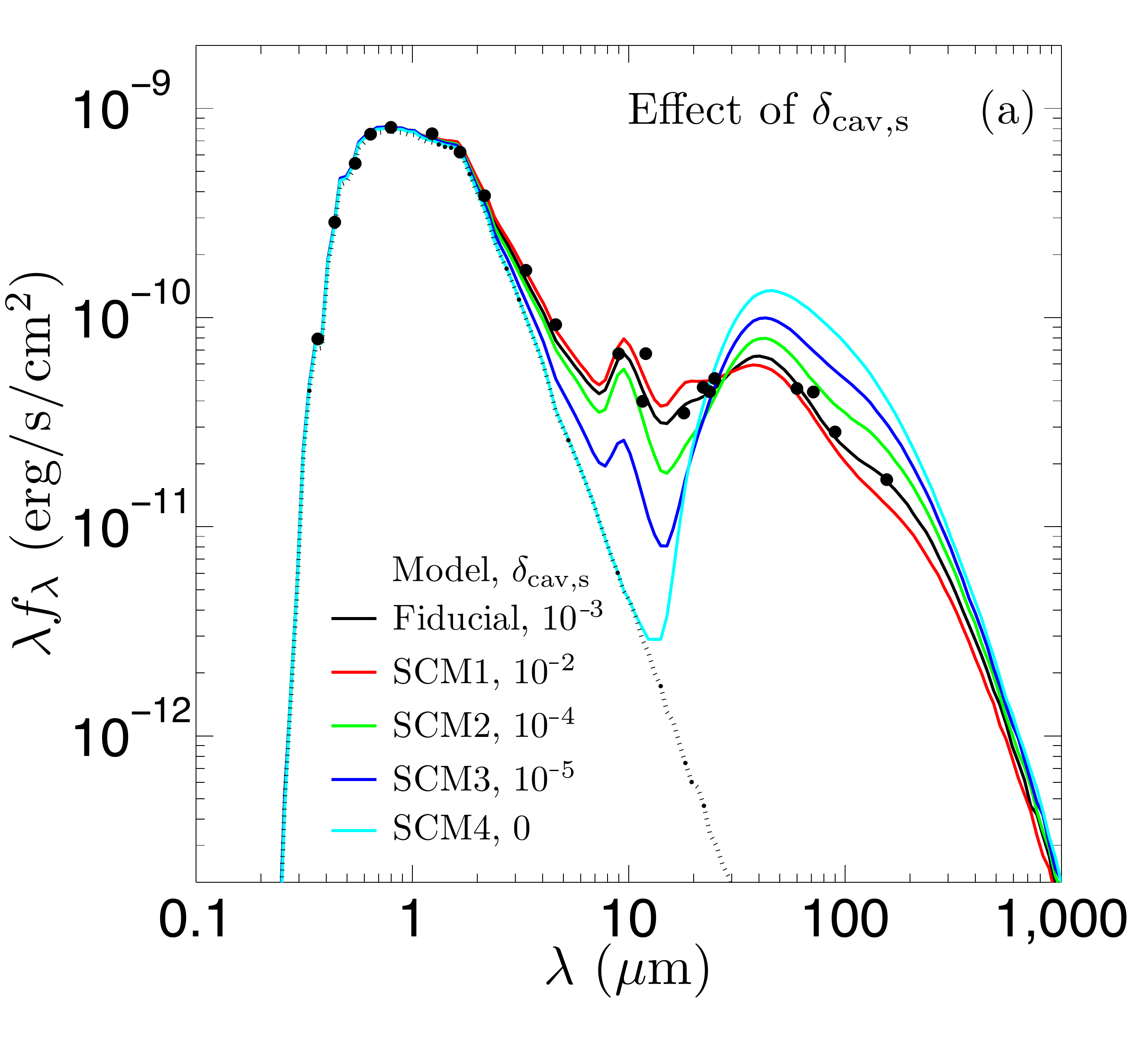} \plotone{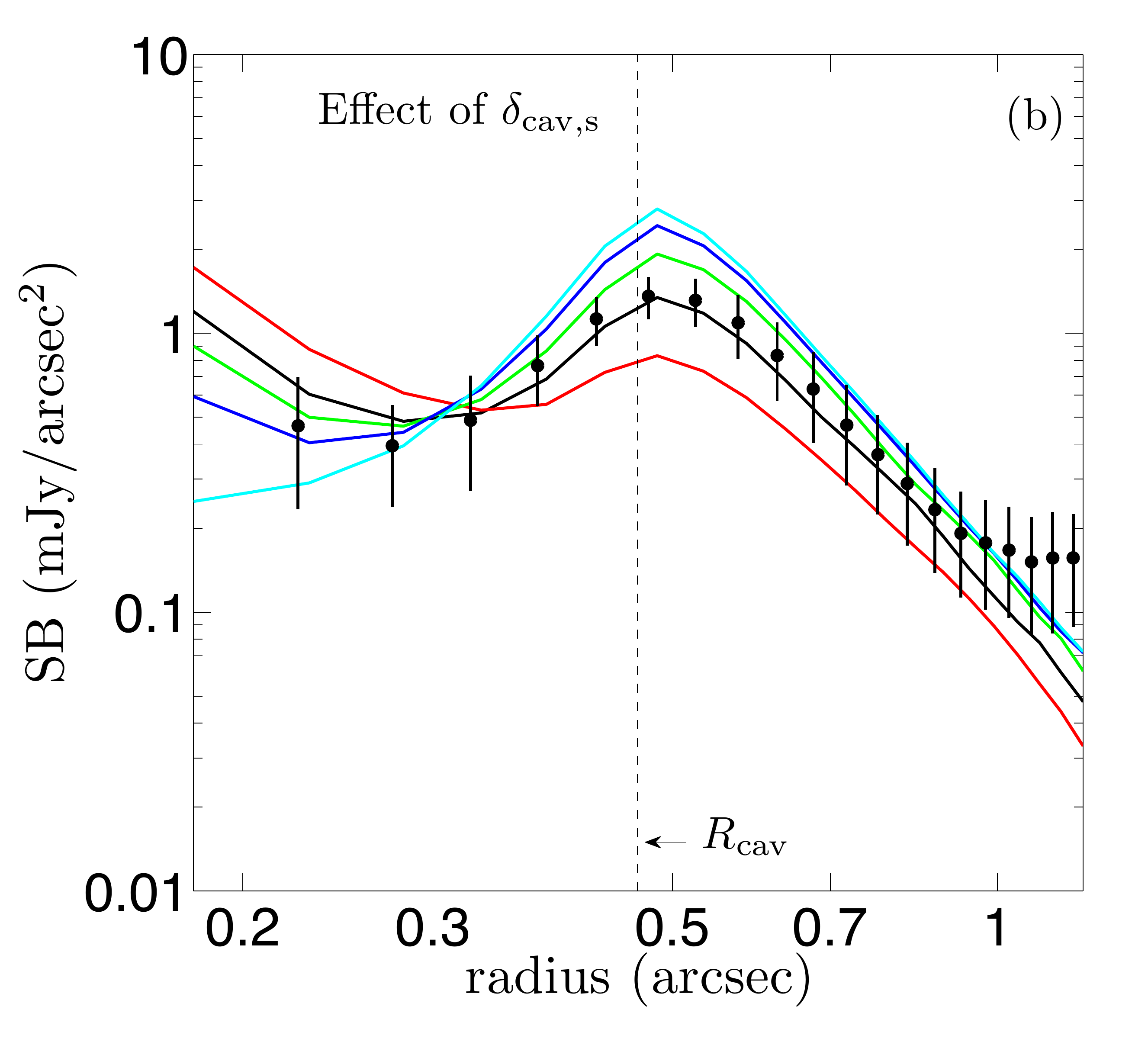} \plotone{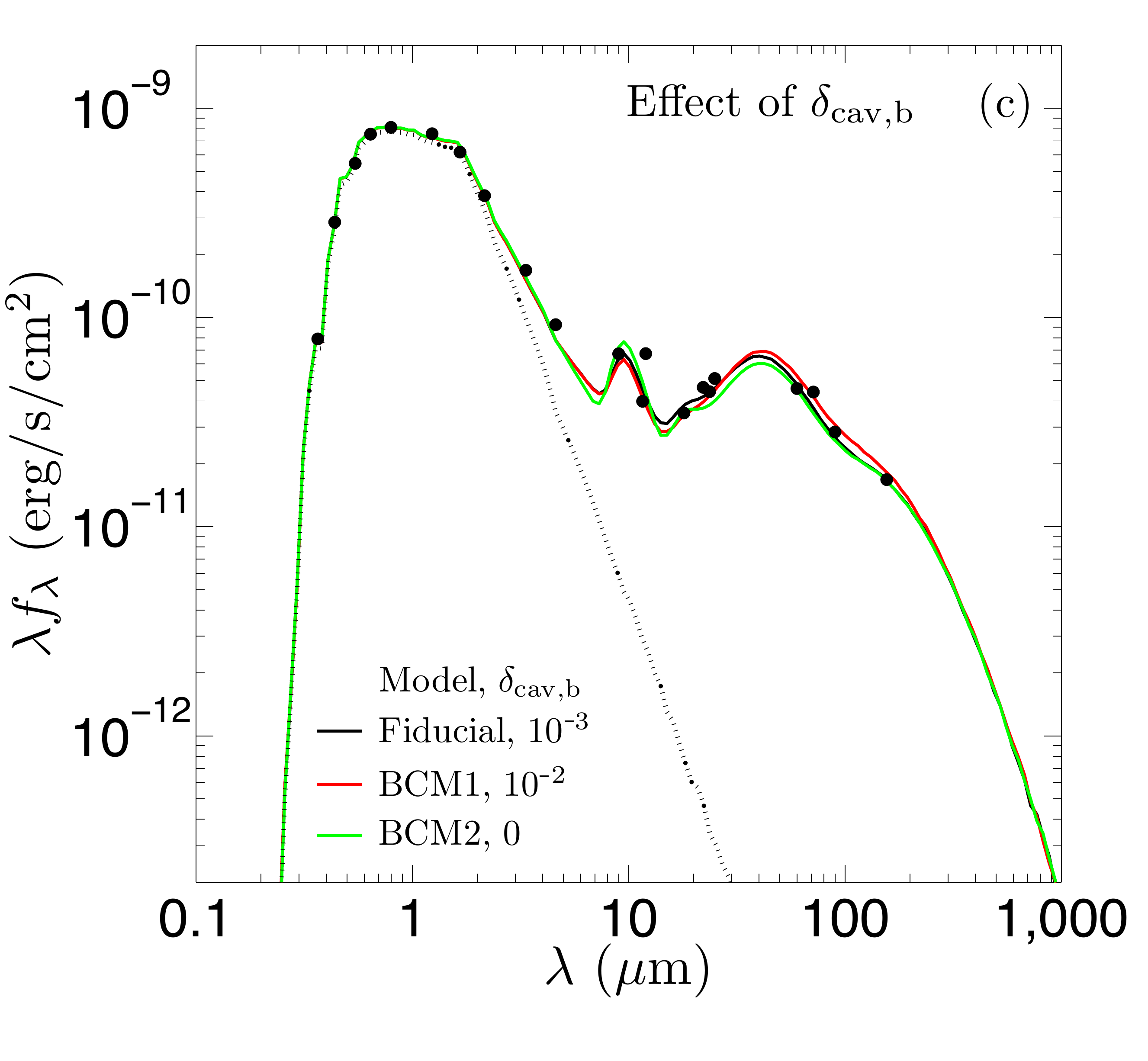} \plotone{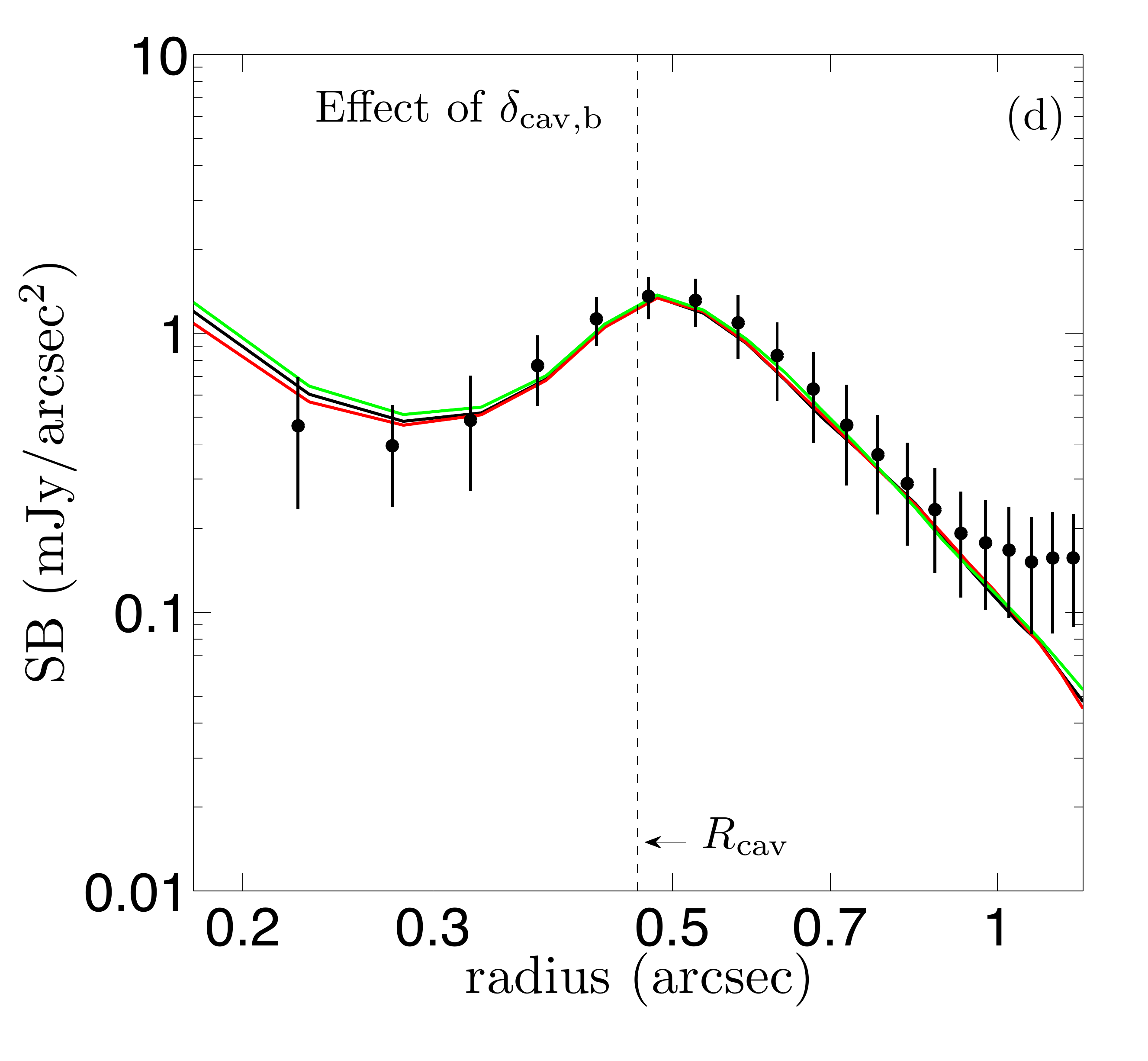}
\end{center}
\figcaption{Effects of the cavity depletion factor for the small ($\dcs$) and big ($\dcb$) dust, on both the disk SED ((a) and (c)) and the scattered light image ((b) and (d), plotting SBRP along the major axes). The dotted line in panels (a) and (c) is the stellar spectrum and the black dots are photometric SED data point, as in Figure~\ref{fig:fiducial-sed-sbrp}. The corresponding disk models are listed in the block ``The Depletion Factor Inside the Cavity'' in Table~\ref{tab:model}, and described in detail in Section~\ref{sec:depletion}. The models in the SBRP plots ((b) and (d)) have the same line types as they have in the SED plots ((a) and (c)). The results show that as long as the innermost disk is vertically optically thick, the SED only weakly depends on $\dcs$; once $\dcs$ drops enough for the innermost disk to be optically thin, the NIR excess becomes sensitive to the amount of small dust there. The inner disk is brighter and the outer disk is dimmer in scattered light for a bigger $\dcs$. On the other hand, both the SED and scattered light image are almost independent of $\dcb$.
\label{fig:depletion}}
\end{figure}

\begin{figure}[tb]
\begin{center}
\epsscale{0.45} \plotone{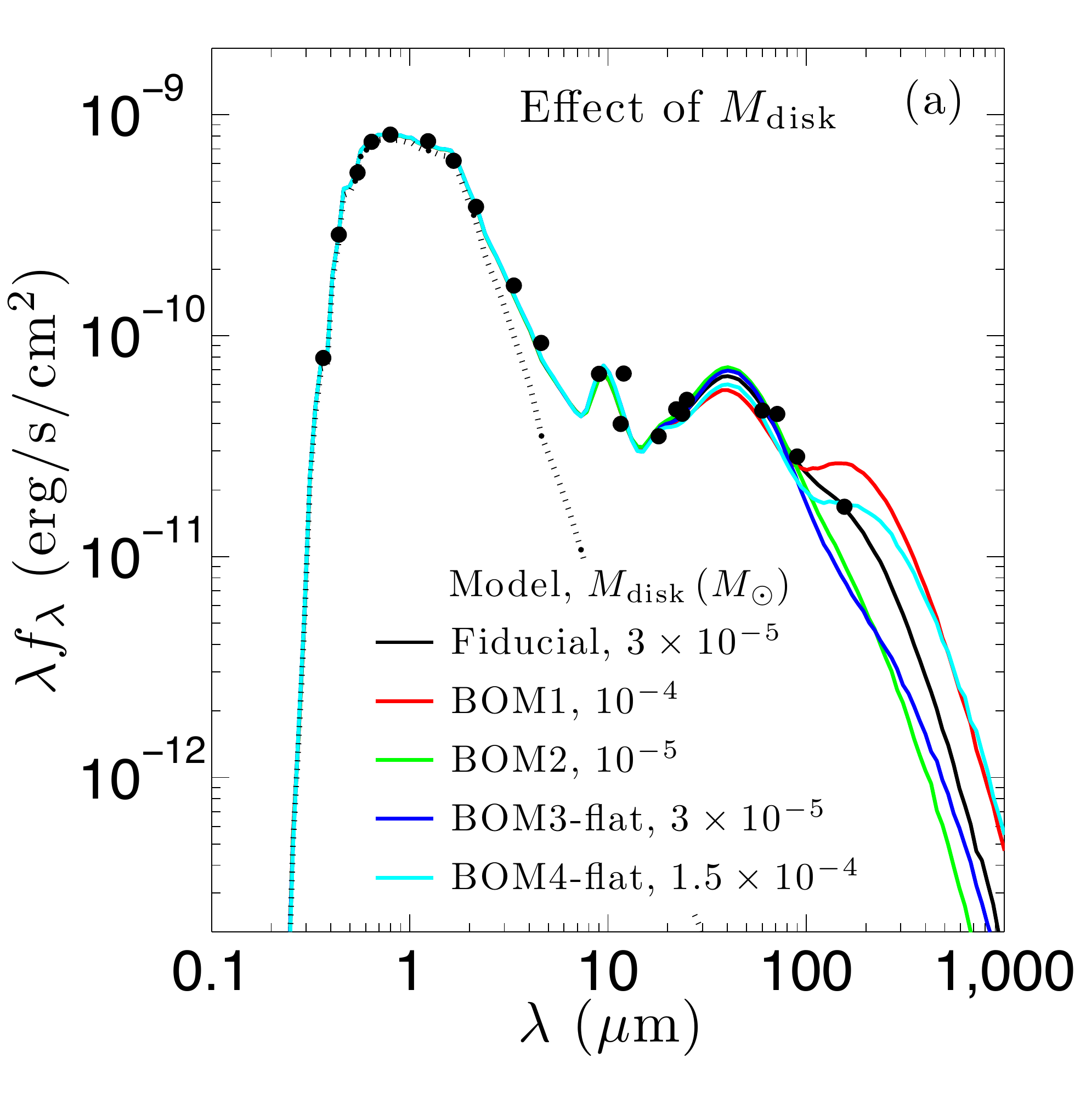} \plotone{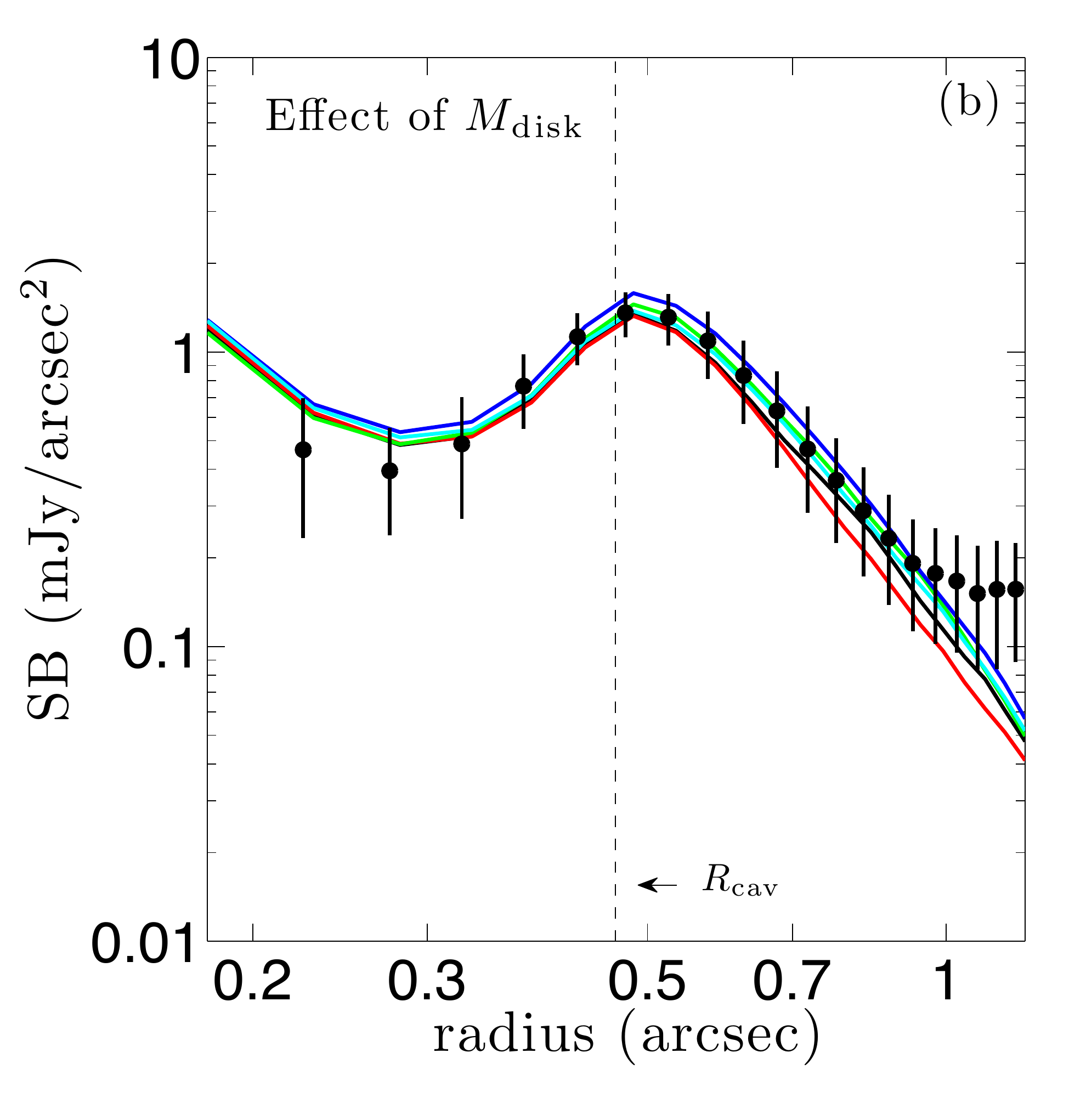}
\end{center}
\figcaption{The same set of plots as in Figure~\ref{fig:depletion}, but for models BOM1 to BOM4-flat, showing the effect of total disk mass $\md$. The corresponding disk models are listed in the block ``The Total Dust Mass of the Disk'' in Table~\ref{tab:model}, and described in detail in Section~\ref{sec:mass}. The dust model for the big dust in models BOM3-flat and BOM4-flat is the flat-big-dust, while the other models have the steep-big-dust model (Figure~\ref{fig:kappa}). The total mass of the disk, which in PDS 70 concentrates in the big dust in the outer disk, needs to be better constrained by observations at long wavelengths, with which different big dust models can be distinguished. The photometry at the longest wavelength available (160$\mu$m) agrees with a total disk mass $\sim3\times10^{-5}M_\odot$ assuming the steep-big-dust model, or $\sim1.5\times10^{-4}M_\odot$ assuming the flat-dust-model.
\label{fig:mass}}
\end{figure}

\begin{figure}[tb]
\begin{center}
\epsscale{0.45} \plotone{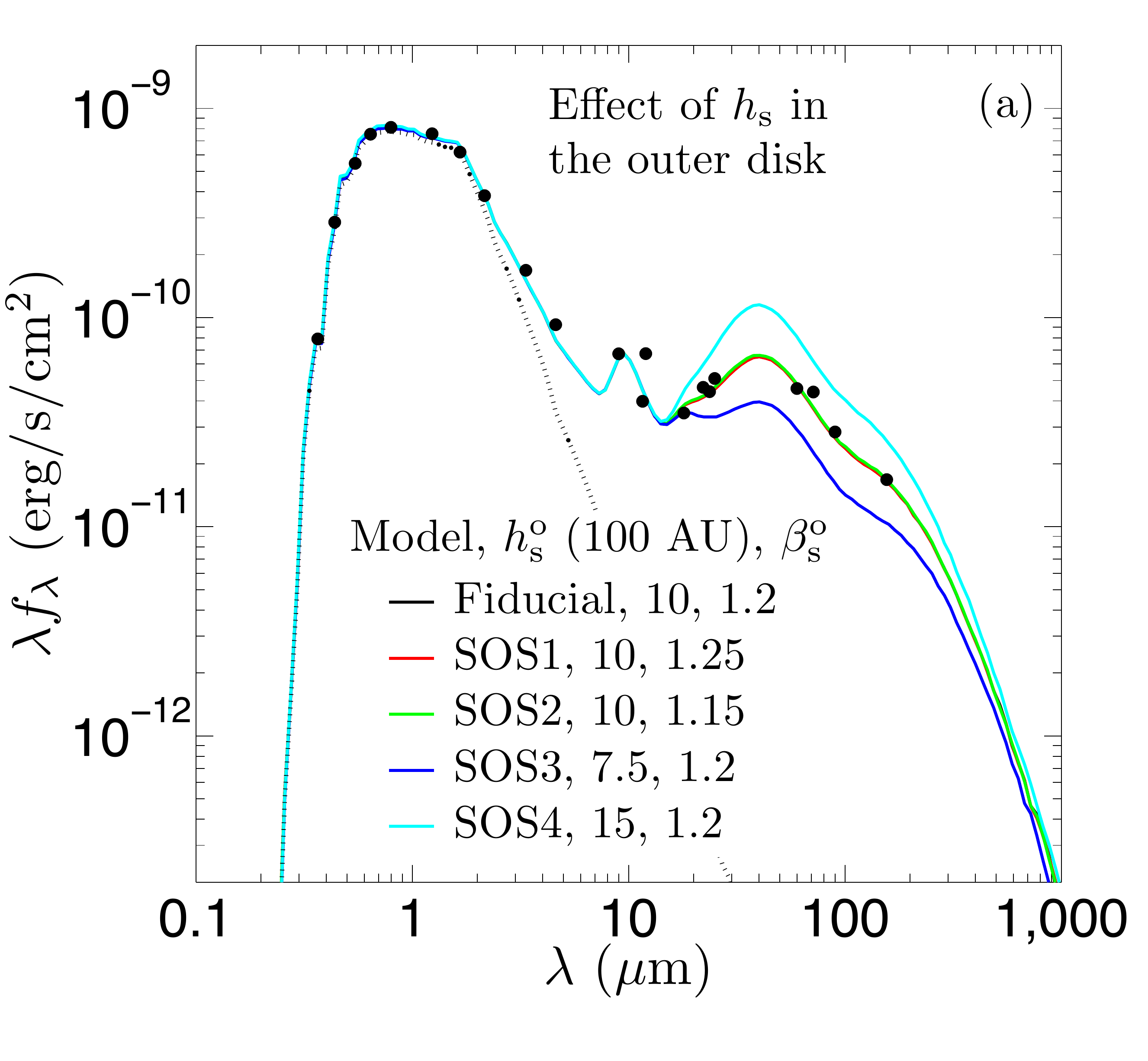} \plotone{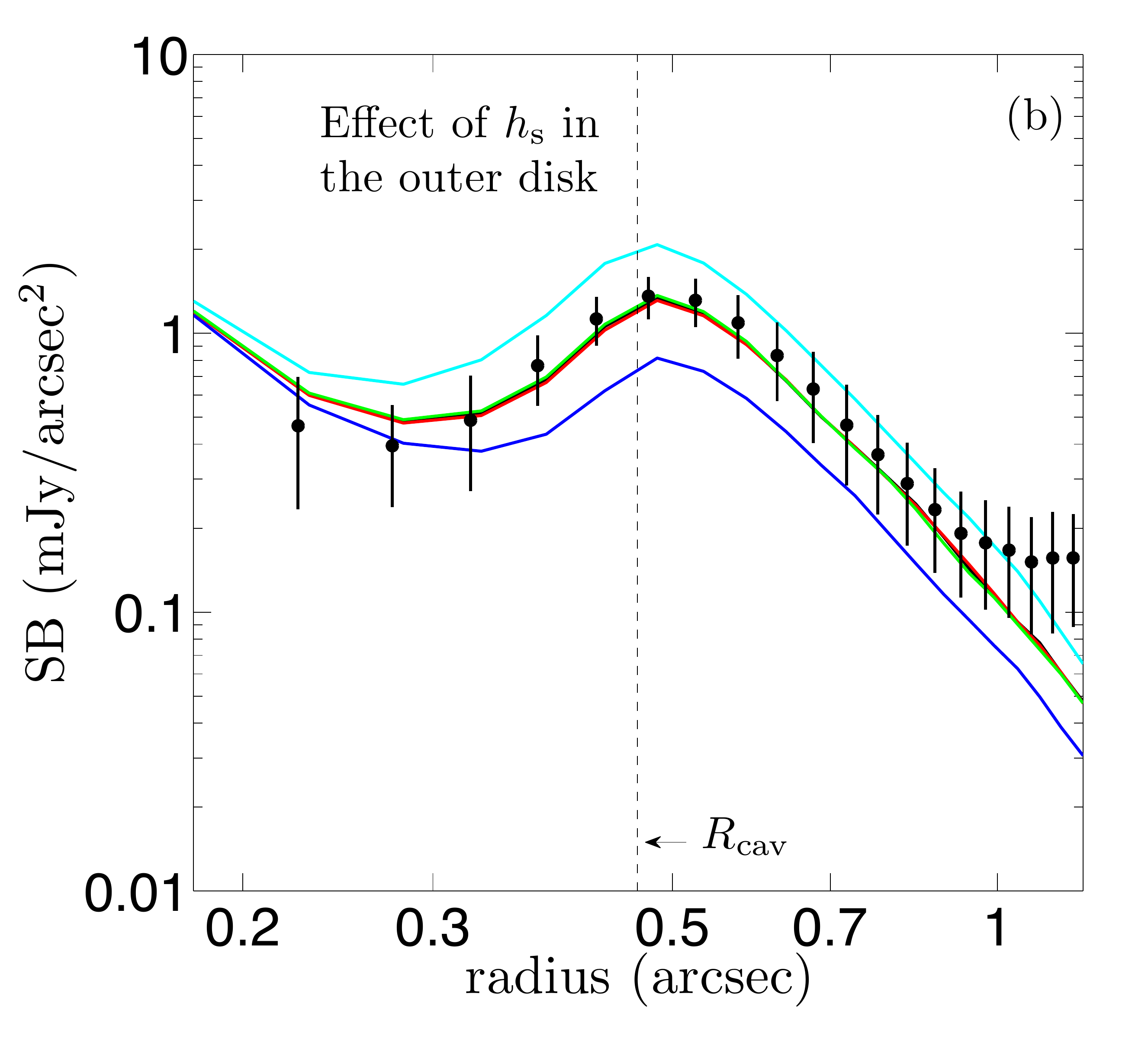} \plotone{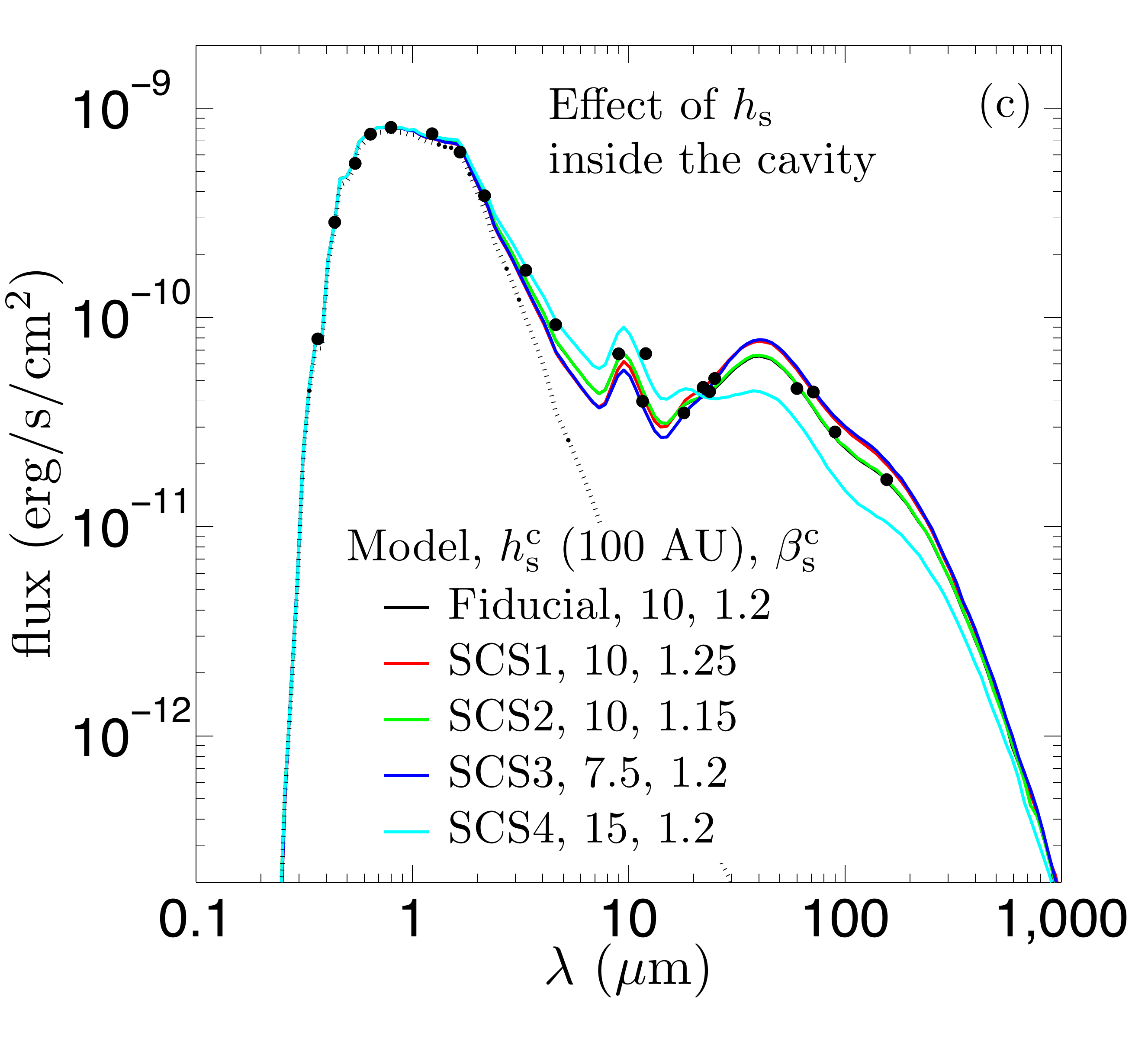} \plotone{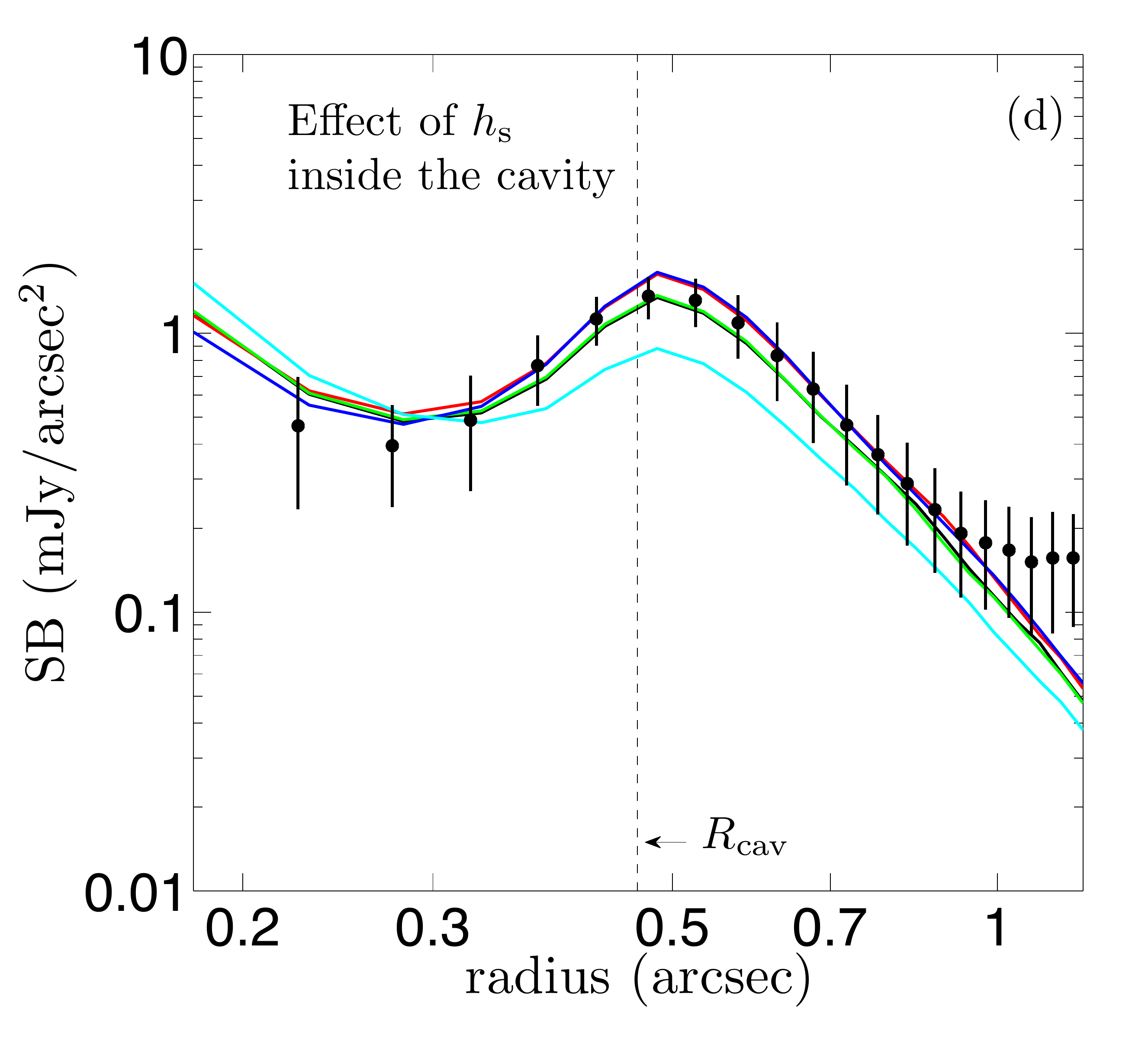}
\end{center}
\figcaption{The same set of plots as in Figure~\ref{fig:depletion}, but for models SOS1-4 and SCS1-4, showing the effects of scale height of the small dust, both outside the cavity ($\hso$, (a) and (b) and inside the cavity ($\hsc$, (c) and (d)). The corresponding disk models are listed in the block ``The Scale Height of the Small Dust'' in Table~\ref{tab:model}, and described in detail in Section~\ref{sec:scaleheight}. Both the IR excess at $\sim40\mu$m and the brightness of the disk at the cavity edge sensitively depend on the overall scale of $\hso$ (a higher cavity wall produces more MIR excess and scattered light), while if the overall scale is roughly the same, the radial dependence of $\hso$ does not have a big effect. Similar results are seen in $\hsc$ as well.
\label{fig:scaleheight}}
\end{figure}